\documentclass[aip,jap,preprint,amsmath,amssymb,floatfix,raggedbottom]{revtex4-2}

\usepackage{natbib}
\usepackage{graphicx}
\usepackage{placeins}
\usepackage{mathtools}
\usepackage{booktabs}
\usepackage{siunitx}
\usepackage{xcolor}
\usepackage{microtype}
\usepackage{tikz}
\usetikzlibrary{arrows.meta,calc,positioning,shapes.geometric}
\usepackage[colorlinks=true,allcolors=blue!55!black]{hyperref}
\usepackage[nameinlink,noabbrev]{cleveref}

\graphicspath{{figures/}}

\newcommand{\WmK}{\unit{\watt\per\meter\per\kelvin}}

\newcommand{\K}{\mathbf K}

\newcommand{\Kf}{\mathbf K_{f}}

\newcommand{\A}{\boldsymbol{\Lambda}}
\newcommand{\bvec}{\boldsymbol{\mu}}
\newcommand{\Smat}{\mathbf S}

\newcommand{\nuv}{\boldsymbol\nu}

\newcommand{\Pmat}{\mathbf P}
\newcommand{\Jmat}{\mathbf J}

\newcommand{\iu}{\mathrm i}
\newcommand{\dd}{\mathrm d}

\newcommand{\diag}{\operatorname{diag}}

\begin{document}

\title{\texorpdfstring{%
Identifiability of the three-dimensional anisotropic\\
thermal-conductivity tensor in thermoreflectance measurements:\\
Single-orientation limits and full-tensor recovery%
}{Identifiability of the three-dimensional anisotropic thermal-conductivity
tensor in thermoreflectance measurements: Single-orientation limits and
full-tensor recovery}}

\author{Dihui Wang}
\affiliation{Department of Mechanical Engineering and Materials Science,
University of Pittsburgh, Pittsburgh, PA, USA}
\author{Heng Ban}
\affiliation{Department of Mechanical Engineering and Materials Science,
University of Pittsburgh, Pittsburgh, PA, USA}

\date{11 September 2026}

\begin{abstract}
Materials with full anisotropic thermal conductivity tensors are
widely used in advanced applications,
creating a need for dedicated characterization.
Such tensors contain six independent components: three
diagonal conductivities and three non-zero off-diagonal
couplings. However, whether
thermoreflectance measurements on a single crystalline orientation
can uniquely determine an arbitrary tensor
remains unclear.
Here, we show that every fully anisotropic
thermal-conductivity tensor admits a Schur-equivalent tensor with
zero cross-plane couplings ($k_{xz}=k_{yz}=0$).
   The two tensors produce
identical surface thermal responses, and this equivalence persists as
frequency, delay time, beam geometry, and spatial offset vary.
We further demonstrate that parameter-identifiability assessments based
on heuristic interpretations of sensitivity trends or singular-value
decomposition (SVD) of the sensitivity matrix cannot, by themselves,
rule out globally equivalent solutions in fully anisotropic systems.
Combining systematic
analysis with numerical validation, we provide guidance on selecting
signals and measurement orientations for full-tensor characterization.
Specifically, measurements at three mutually orthogonal crystallographic
orientations provide complementary information for full-tensor recovery.
These findings clarify the limits of single-surface measurements and
establish a systematic approach to characterizing fully anisotropic
thermal-conductivity tensors.
\end{abstract}

\maketitle

\clubpenalty=10000
\widowpenalty=10000

\providecommand{\IntroductionFile}{sections/01_introduction}

\section{Introduction}
\label{sec:introduction}
Macroscopic heat conduction in an anisotropic solid obeys the tensor form of
Fourier's law, $\mathbf q=-\K\nabla T$, where $\mathbf q$ is the heat-flux
vector, $\nabla T$ is the temperature gradient, and $\K$ is the symmetric
second-rank thermal-conductivity tensor.  In a right-handed Cartesian frame
$(\hat{\mathbf x},\hat{\mathbf y},\hat{\mathbf z})$, the tensor in matrix form
is
\begin{equation}
  \K=
  \begin{bmatrix}
    k_{xx} & k_{xy} & k_{xz}\\
    k_{xy} & k_{yy} & k_{yz}\\
    k_{xz} & k_{yz} & k_{zz}
  \end{bmatrix}.
  \label{eq:tensor}
\end{equation}
The components of $\K$ reflect both the material and the coordinate frame in which they
are expressed. A full anisotropic thermal-conductivity tensor
refers to the most general form, in which all six components are
independent: the diagonal terms $k_{xx}$, $k_{yy}$, and $k_{zz}$, and the
off-diagonal terms $k_{xy}$, $k_{xz}$, and $k_{yz}$.

Materials with fully anisotropic thermal conductivity tensors play two
distinct roles in advanced applications. First, a material may be selected
primarily for its electronic, optical, or structural functionality, while its
thermal anisotropy imposes a constraint that must be accommodated in the
thermal design. Representative examples include the wide-bandgap power
semiconductor \(\beta\)-Ga\(_2\)O\(_3\)
\citep{Jiang2018Ga2O3,Klimm2023Ga2O3}, laser-host crystals such as
KLu(WO\(_4\))\(_2\) \citep{Silvestre2008KLuWO4}, and environmental-barrier
coatings such as \(\beta\)-Y\(_2\)Si\(_2\)O\(_7\)
\citep{Olson2020Y2Si2O7}.
Second, anisotropic thermal transport may itself provide the engineered
functionality. Thermal metamaterials are designed with anisotropic conductivity
to guide, concentrate, or cloak heat flow
\citep{Li2021ThermalMetamaterials,Narayana2012HeatFlux,Vemuri2014Guiding}.
Artificially tilted thermoelectric multilayers extend the
rotated-laminate heat-routing mechanism to energy conversion, producing a
transverse voltage from a longitudinal temperature difference
\citep{Vemuri2014Guiding,Ando_OffDiagonalConduction}.
In both cases, device performance and reliability depend on anisotropic
three-dimensional heat transport, and the underlying conductivity tensor must
therefore be characterized accurately.

Non-contact thermoreflectance and lock-in thermography provide several routes
to measure anisotropic conductivity. Time-domain thermoreflectance (TDTR)
and frequency-domain thermoreflectance (FDTR) vary delay, modulation
frequency, pump--probe offset, or beam ellipticity to probe in-plane
anisotropy~\cite{Feser2012,Jiang2018,Tang2021}. Spatial-domain
thermoreflectance (SDTR) uses spatial phase scans~\cite{Jiang2022SpatialDomain},
whereas spatially resolved lock-in micro-thermography (SR-LIT) fits
two-dimensional maps through tensor analysis~\cite{Wang2024}.
Complementary iterative transient electrothermal fitting characterizes
the thermal properties of thin fibers~\cite{Spirnock2026Fibers}.

\begin{figure}[!t]
  \centering
  \includegraphics[width=0.98\textwidth]{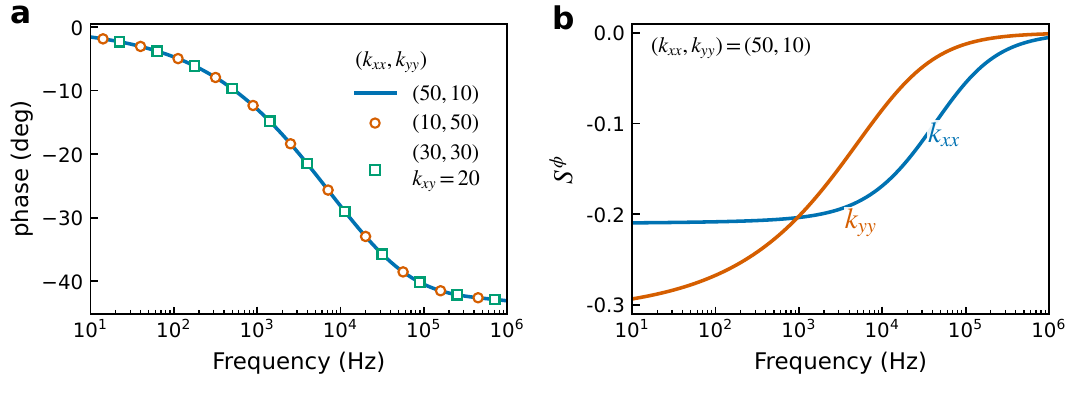}
  \caption{Azimuthal degeneracy and phase sensitivity of centered
circular FDTR on a substrate with a known isotropic metal transducer
(coating parameters in Supplementary Note~S6). (a) FDTR phase versus frequency for
$(k_{xx},k_{yy})=(50,10)$, $(10,50)$, and $(30,30)$ $(\WmK)$; the third case
additionally has $k_{xy}=20\,\WmK$. All three tensors have
$k_{zz}=10\,\WmK$ and $k_{xz}=k_{yz}=0$.
They share the in-plane eigenvalue pair $(50,10)$ and differ only by rotation
about $\hat z$. The reference curve (solid) and
rotated-tensor symbols coincide to machine precision.
    (b) Dimensionless normalized phase sensitivities $S_{p}^{\phi}$ for
$p=k_{xx}$ and $k_{yy}$, evaluated at $(k_{xx},k_{yy})=(50,10)\,\WmK$.
The distinct local sensitivities do not remove the global rotation ambiguity.}
  \label{fig:inplane-rotation-ambiguity}
\end{figure}

Directional approaches resolve in-plane anisotropy
($k_{xy}\neq0$ and $k_{xz}=k_{yz}=0$) by producing nearly one-dimensional
heat flow along different in-plane directions. The resulting directional
conductivities are then combined to reconstruct the in-plane tensor.
For three-dimensional heat flow in polyimide films, square-pulsed-source
thermoreflectance resolves in-plane and cross-plane conductivities and
volumetric heat capacity~\cite{Zhang2026Polyimide}.

Tensors with nonzero cross-plane couplings $k_{xz}$ and $k_{yz}$ present a
more difficult inverse problem. One route combines measurements in known
crystal orientations. Olson \emph{et al.} used cross-plane TDTR on
EBSD-indexed grains in polycrystalline $\beta$-Y$_2$Si$_2$O$_7$
~\citep{Olson2020Y2Si2O7}; this requires sufficiently diverse orientations
and grains large compared with the thermal footprint. Three-dimensional
SR-LIT instead used three orthogonal crystal orientations~\citep{Wang2025}.
A second route uses a constrained material model: BO-SPS reported all six
components for AT-cut quartz from beam offsets on one
orientation~\citep{Chen2025}. Its transfer matrix, however, propagates the
incomplete cross-plane heat flux $Q=-k_{zz}\partial_z T$. As shown below,
omitting the in-plane contributions to the complete flux
$q_z=-(k_{xz}\partial_x T+k_{yz}\partial_y T+k_{zz}\partial_z T)$ creates
artificial offset asymmetry. Whether a single-orientation measurement can
uniquely determine an unconstrained tensor therefore requires a direct
identifiability analysis.

%


Multiparameter fitting is commonly assessed through sensitivity analysis.
For an observable $g(\chi)$, the sensitivity to a parameter $p_j$ is
$S_{p_j}^{g}(\chi)=\partial\ln|g(\chi)|/\partial\ln|p_j|
=(p_j/g)\,\partial g(\chi)/\partial p_j$;
$\chi$ denotes the controlled experimental condition, such as modulation
frequency, delay time, or spatial position.
Schmidt \emph{et al.} proposed using separated sensitivity trends to assess
the feasibility of multiparameter fitting~\cite{Schmidt2009}. Such criteria
guide experimental design but do not prove uniqueness.
More recently, sensitivity-matrix reduction and singular-value decomposition
(SVD) have provided a systematic test of the local linear independence of a
specified set of parameters~\cite{TuOng2023}.
Chen et al. proposed that the number of parameters that can be determined is
the number of singular values satisfying the practical cutoff
$\sigma_i>0.1$~\cite{ChenJiang2025}.

Neither sensitivity analysis nor SVD, however, establishes global uniqueness.
In the centered circular FDTR example of
\cref{fig:inplane-rotation-ambiguity}, the two in-plane conductivities have
distinct sensitivities and both singular values of the unnormalized
log-phase sensitivity matrix exceed the proposed cutoff.
Yet swapping the conductivities leaves the response unchanged. Allowing
$k_{xy}\ne0$ extends this ambiguity to rotated tensors with the same
in-plane eigenvalues: the signal depends only on their trace and determinant.
Thus a locally independent parameter pair need not have a unique laboratory
axis assignment. Supplementary Note~S5 gives the symmetry proof and numerical
spectra.

This work focuses on thermoreflectance measurements of an anisotropic
substrate beneath a metal transducer, as used in standard FDTR and TDTR.
We derive the Schur-equivalent matrix and the resulting single-orientation
information limit for the coated planar stack. We then combine
sensitivity-matrix analysis with global-equivalence checks to guide signal
selection and measurement geometry. Three registered orthogonal surfaces
provide a route to an otherwise unconstrained tensor. We demonstrate this
route with a joint six-entry fit to noisy phase-frequency scans and verify
that the combined sensitivity matrix resolves all six parameter directions.


\section{Three-dimensional anisotropic model and single-orientation theorem}
\label{sec:model}

We derive the harmonic surface response of a planar multilayer
structure in terms of the Schur-equivalent thermal-conductivity tensor.

\subsection{Governing equation and face-local formulation}
\label{sec:geometry-partition}

Within each homogeneous layer, the heat-diffusion equation
$C_v\partial_tT=\nabla\!\cdot(\Kf\nabla T)$ expands to
\begin{equation}
  C_v\frac{\partial T}{\partial t}
  =k_{xx}\frac{\partial^2T}{\partial x^2}
  +k_{yy}\frac{\partial^2T}{\partial y^2}
  +k_{zz}\frac{\partial^2T}{\partial z^2}
  +2k_{xy}\frac{\partial^2T}{\partial x\,\partial y}
  +2k_{xz}\frac{\partial^2T}{\partial x\,\partial z}
  +2k_{yz}\frac{\partial^2T}{\partial y\,\partial z}.
  \label{eq:time-domain-diffusion}
\end{equation}
Here, $T=T(x,y,z,t)$ is the temperature rise relative to equilibrium, $C_v$ is
the volumetric heat capacity, $t$ is time, and $\Kf$ is the
thermal-conductivity tensor of that layer, written in the form of
\cref{eq:tensor}.  The $x$ and $y$
directions are in-plane, while $z$ is the inward cross-plane coordinate.
Heating and detection occur at the outer transducer surface. Neglecting
radiative and convective losses there, we represent
the modulated component of an absorbed continuous-wave laser as
$q_p(x,y,t)=q_{p,0}(x,y)e^{\iu\omega_0t}$, where $\omega_0$ is the
modulation angular frequency.  With the temporal transform convention
$\mathcal F_t\{f\}=\int_{-\infty}^{\infty}f(t)e^{-\iu\omega t}\,\dd t$, the
boundary condition is $q_z(x,y,0;\omega)=q_p(x,y;\omega)$, where
\begin{equation}
  \begin{aligned}
    q_p(x,y;\omega)
    &=\frac{2P_{\mathrm{abs}}}{\pi w_xw_y}
      \exp\!\left[-2\left(\frac{(x-x_p)^2}{w_x^2}
                    +\frac{(y-y_p)^2}{w_y^2}\right)\right]\\
    &\quad\times2\pi\delta(\omega-\omega_0).
  \end{aligned}
  \label{eq:gaussian-pump}
\end{equation}
in which $P_{\mathrm{abs}}$ is the absorbed modulated-power amplitude,
$(x_p,y_p)$ the pump center, and $w_x$ and $w_y$ the $1/e^2$ intensity radii.
The factor $2\pi\delta(\omega-\omega_0)$ selects the modulation frequency and
is suppressed in the phasor equations below, evaluated at $\omega=\omega_0$.
We now derive the surface response using the Schur-equivalent matrix;
Supplementary Note~S1 gives the Fourier-transform details.

\subsection{Solution based on the Schur-equivalent matrix}

We partition the thermal-conductivity tensor as
\begin{equation}
  \begin{aligned}
    \Kf&=
    \begin{bmatrix}
      \A & \bvec\\
      \bvec^{\mathsf T} & k_{zz}
    \end{bmatrix}\in\mathbb R^{3\times3},\\
    \A&=\begin{bmatrix}k_{xx}&k_{xy}\\k_{xy}&k_{yy}\end{bmatrix}
      \in\mathbb R^{2\times2},
    &\bvec&=\begin{bmatrix}k_{xz}\\k_{yz}\end{bmatrix}
      \in\mathbb R^{2\times1},
  \end{aligned}
  \label{eq:face-partition}
\end{equation}
where $\A$ is the in-plane block, $\bvec$ couples the in-plane and cross-plane
directions, and $k_{zz}$ is the cross-plane conductivity.
We define the $2\times2$ Schur complement
\begin{equation}
  \Smat=\A-\frac{\bvec\bvec^{\mathsf T}}{k_{zz}}.
  \label{eq:schur-definition}
\end{equation}

Taking the Fourier transform with respect to the in-plane coordinates
$(x,y)$ and time $t$, we write
\begin{equation}
  T(x,y,z,t)
  \xleftrightarrow{\;\mathcal{F}_{x,y,t}\;}
  \widehat{\theta}(u,v,z;\omega),
\end{equation}
where $u$ and $v$ are the in-plane spatial frequencies and $\omega$ is the
angular frequency. With the Fourier-space gradient operator
$\widehat\nabla=(\iu2\pi u,\iu2\pi v,\partial_z)^{\mathsf T}$,
applying the transform to \cref{eq:time-domain-diffusion} and Fourier's law gives
\begin{equation}
  \iu\omega C_v\widehat\theta
  =\widehat\nabla^{\mathsf T}\Kf\widehat\nabla\widehat\theta,
  \qquad \widehat{\mathbf q}=-\Kf\widehat\nabla\widehat\theta,
  \label{eq:harmonic-pde}
\end{equation}
where $\widehat\theta$ and $\widehat{\mathbf q}$ are the complex
Fourier-space temperature and heat-flux amplitudes. The in-plane gradient
transforms to $\widehat\nabla_t\widehat\theta
=\iu2\pi(u,v)^{\mathsf T}\widehat\theta$, so the cross-plane heat flux is
\begin{equation}
  \widehat q_z=-\bvec^{\mathsf T}\widehat\nabla_t\widehat\theta
  -k_{zz}\,\partial_z\widehat\theta .
  \label{eq:physical-normal-flux}
\end{equation}

Let column vector $\nuv\equiv[u\;v]^{\mathsf T}\in\mathbb R^{2\times1}$, so
that $\nabla_t\mapsto\iu2\pi\nuv$.  Supplementary Note~S1 gives the transform
pair and the term-by-term reduction.  At fixed $(\nuv,\omega)$, primes denote
derivatives with respect to depth.  We define the shear coefficient
\begin{equation}
  \beta(\nuv)=2\pi\nuv^{\mathsf T}\bvec=2\pi(uk_{xz}+vk_{yz}).
  \label{eq:beta-definition}
\end{equation}
Then \cref{eq:harmonic-pde} becomes
\begin{equation}
  k_{zz}\widehat\theta''+2\iu\beta\,\widehat\theta'
  -\left[(2\pi)^2\nuv^{\mathsf T}\A\nuv+\iu\omega C_v\right]
  \widehat\theta=0 .
  \label{eq:fourier-depth-ode}
\end{equation}

All explicit $\bvec$ dependence in \cref{eq:fourier-depth-ode} resides in the
first-derivative term, and within a homogeneous layer $\beta$ and $k_{zz}$ are
independent of $z$.  Division by $k_{zz}$ places the equation in the form
$\widehat\theta''+P\widehat\theta'+Q\widehat\theta=0$ with
\begin{equation}
  P=\frac{2\iu\beta}{k_{zz}},\qquad
  Q=-\frac{(2\pi)^2\nuv^{\mathsf T}\A\nuv+\iu\omega C_v}{k_{zz}},
  \label{eq:normal-form-coefficients}
\end{equation}
so the standard reduction to normal form,
$\widehat\theta=\exp[-\tfrac12\int P\,\dd z]\,A$, becomes the unimodular
substitution
\begin{equation}
  \widehat\theta(z)=\exp\!\left(-\frac{\iu\beta z}{k_{zz}}\right)A(z).
  \label{eq:gauge-substitution}
\end{equation}
This substitution separates the shear phase from the transformed temperature
amplitude $A(z)$. Its derivatives are
\begin{align*}
  \widehat\theta'
  &=e^{-\iu\beta z/k_{zz}}
    \left(A'-\frac{\iu\beta}{k_{zz}}A\right),\\
  \widehat\theta''
  &=e^{-\iu\beta z/k_{zz}}
    \left(A''-\frac{2\iu\beta}{k_{zz}}A'
    -\frac{\beta^2}{k_{zz}^2}A\right).
\end{align*}
Upon substitution into \cref{eq:fourier-depth-ode}, the contributions
$-2\iu\beta A'$ and $+2\iu\beta A'$ cancel.  Removing the common exponential
factor gives
\begin{equation}
  k_{zz}A''-
  \left[(2\pi)^2\nuv^{\mathsf T}\A\nuv
  -\frac{\beta^2}{k_{zz}}+\iu\omega C_v\right]A=0 .
  \label{eq:transformed-depth-intermediate}
\end{equation}
Using $\beta=2\pi\nuv^{\mathsf T}\bvec$ and
$\Smat=\A-\bvec\bvec^{\mathsf T}/k_{zz}$, we obtain

\begin{equation}
  (2\pi)^2\nuv^{\mathsf T}\A\nuv-\frac{\beta^2}{k_{zz}}
  =(2\pi)^2\nuv^{\mathsf T}\Smat\nuv,
  \label{eq:schur-quadratic-identity}
\end{equation}

so that \cref{eq:transformed-depth-intermediate} becomes
\begin{equation}
  k_{zz}A''
  -\left[(2\pi)^2\nuv^{\mathsf T}\Smat\nuv+\iu\omega C_v\right]A=0 .
  \label{eq:reduced-ode}
\end{equation}
Equation~\eqref{eq:reduced-ode} is the governing equation associated with the
Schur-equivalent matrix $\diag(\Smat,k_{zz})$.
The propagation constants of \cref{eq:reduced-ode} are $\pm\sigma$
\begin{equation}
  \sigma=\sqrt{\frac{\iu\omega C_v+(2\pi)^2
  \nuv^{\mathsf T}\Smat\nuv}{k_{zz}}},
  \qquad \operatorname{Re}\sigma>0,
  \label{eq:decay-constant}
\end{equation}
The corresponding modes of \cref{eq:fourier-depth-ode} are
\begin{equation}
  \widehat\theta=A_+e^{s_+z}+A_-e^{s_-z},
  \qquad
  s_{\pm}=-\frac{\iu\beta}{k_{zz}}\pm\sigma ,
  \label{eq:depth-roots}
\end{equation}
in which $s_+$ and $s_-$ are the two complex depth roots and $A_+$ and $A_-$
are their complex temperature amplitudes. The shear contributes an imaginary
offset, while $\sigma$ sets attenuation and temporal phase lag and depends on
$\bvec$ only through $\Smat$.

The cross-plane heat flux in \cref{eq:physical-normal-flux} transforms
covariantly under the same shear:
\begin{equation}
  \widehat q_z
  =-\left(k_{zz}\frac{\dd}{\dd z}+\iu\beta\right)\widehat\theta
  =-e^{-\iu\beta z/k_{zz}}\,k_{zz}A' ,
  \label{eq:transformed-full-flux}
\end{equation}
The temperature and heat flux form the state vector
\begin{equation}
  \begin{bmatrix}
    \widehat\theta(z)\\
    \widehat q_z(z)
  \end{bmatrix}
  =
  \begin{bmatrix}
    e^{s_+z} & e^{s_-z}\\
    -k_{zz}\sigma e^{s_+z} & k_{zz}\sigma e^{s_-z}
  \end{bmatrix}
  \begin{bmatrix}
    A_+\\
    A_-
  \end{bmatrix}.
  \label{eq:modal-state-matrix}
\end{equation}
Eliminating the modal amplitudes using the state at $z_0$ gives
\begin{equation}
  \begin{aligned}
  \begin{bmatrix}
    \widehat\theta(z_0+d)\\
    \widehat q_z(z_0+d)
  \end{bmatrix}
  &=e^{-\iu\beta d/k_{zz}}
  \begin{bmatrix}
    \cosh(\sigma d) &
    -\dfrac{\sinh(\sigma d)}{k_{zz}\sigma}\\[6pt]
    -k_{zz}\sigma\sinh(\sigma d) &
    \cosh(\sigma d)
  \end{bmatrix}\\
  &\quad\times
  \begin{bmatrix}
    \widehat\theta(z_0)\\
    \widehat q_z(z_0)
  \end{bmatrix}.
  \end{aligned}
  \label{eq:physical-transfer-matrix}
\end{equation}
The explicit coupling dependence appears only in the common phase factor
$e^{-\iu\beta d/k_{zz}}$, which cancels from the temperature-to-flux ratio. The
two columns of \cref{eq:modal-state-matrix} likewise give the modal admittances
$\widehat q_{z,\pm}/\widehat\theta_\pm=\mp k_{zz}\sigma$.  This cancellation
relies on the in-plane-gradient contribution in
\cref{eq:physical-normal-flux}.

\subsection{Reduced-flux comparator}
\label{sec:reduced-flux-model}

The cancellation in \cref{eq:physical-transfer-matrix} is a property of
the complete cross-plane heat flux, and it is lost if a different boundary variable
is propagated.  Suppose instead that the propagated flux variable is taken to be
$\widehat q_{\mathrm{red}}=-k_{zz}\,\partial_z\widehat\theta$,
which omits the $-\bvec^{\mathsf T}\widehat\nabla_t\widehat\theta$ term.  By
\cref{eq:gauge-substitution},
$\widehat q_{\mathrm{red}}=e^{-\iu\beta z/k_{zz}}
\left(\iu\beta A-k_{zz}A'\right)$.
The common exponential still cancels from
$\widehat q_{\mathrm{red}}/\widehat\theta$, but the explicit $\iu\beta$ term
does not.  For the decaying root,
\begin{equation}
  \frac{\widehat q_{\mathrm{red}}}{\widehat\theta}
  =k_{zz}\sigma+\iu\beta .
  \label{eq:reduced-admittance}
\end{equation}
The surviving term $\iu\beta$ is odd under $\nuv\rightarrow-\nuv$, so for
mirror-symmetric pump and detector weightings the reduced-flux model can
predict unequal phases at opposite offsets. This model-induced asymmetry is
an implementation diagnostic, not evidence that a single-orientation surface
measurement resolves $\bvec$.

\subsection{Surface equivalence for multilayer}
\label{sec:multilayer-theorem}

We number the $N\ge2$ planar layers $j=1,\ldots,N$ from the measured
transducer surface toward the substrate rear. The standard model contains
a metal transducer and an anisotropic substrate separated by a scalar
thermal boundary conductance; additional planar layers are treated identically.  All layers share the same inward $+z$ direction, while layer $j$
carries a local depth coordinate $z_j\in[0,h_j]$ that resets to zero at its
top, in which $h_j$ is the layer thickness.  Denote its volumetric heat
capacity, cross-plane conductivity, coupling vector, and Schur complement by
$C_{v,j}$, $k_{zz,j}$, $\bvec_j$, and $\Smat_j$. The layer-dependent shear
coefficient and propagation constant are
\begin{equation*}
  \begin{aligned}
    \beta_j&=2\pi\nuv^{\mathsf T}\bvec_j,\\
    \sigma_j&=\sqrt{\frac{\iu\omega C_{v,j}+(2\pi)^2
    \nuv^{\mathsf T}\Smat_j\nuv}{k_{zz,j}}},
  \end{aligned}
\end{equation*}
Define the dimensionless thermal thickness $a_j$, characteristic thermal
admittance $Y_j$, and shear factor $p_j$ as
\begin{equation*}
  a_j=\sigma_jh_j,
  \qquad
  Y_j=k_{zz,j}\sigma_j,
  \qquad
  p_j=\exp\!\left(-\iu\frac{\beta_jh_j}{k_{zz,j}}\right).
\end{equation*}
Then \cref{eq:physical-transfer-matrix} becomes

\begin{equation}
  \begin{aligned}
    \begin{bmatrix}
      \widehat\theta_j(h_j)\\
      \widehat q_{z,j}(h_j)
    \end{bmatrix}
    &=\mathbf L_j
    \begin{bmatrix}
      \widehat\theta_j(0)\\
      \widehat q_{z,j}(0)
    \end{bmatrix},\\[4pt]
    \mathbf L_j=p_j\mathbf L_{0,j}
    &=p_j
    \begin{bmatrix}
      \cosh a_j&-\dfrac{\sinh a_j}{Y_j}\\[6pt]
      -Y_j\sinh a_j&\cosh a_j
    \end{bmatrix}.
  \end{aligned}
  \label{eq:layer-factorization}
\end{equation}
The matrix $\mathbf L_{0,j}$ therefore depends on the tensor only through
$(k_{zz,j},\Smat_j)$, while all explicit dependence on
$\beta_j=2\pi\nuv^{\mathsf T}\bvec_j$ is contained in $p_j$.
Supplementary Note~S1 gives the modal derivation of
\cref{eq:layer-factorization}.

An interface with thermal boundary conductance $G_j$ and resistance
$R_j=G_j^{-1}$ preserves the cross-plane heat flux and imposes the temperature
jump $\widehat\theta_{j+1}(0)=\widehat\theta_j(h_j)
-R_j\widehat q_{z,j}(h_j)$:
\begin{equation}
  \begin{aligned}
    \begin{bmatrix}
      \widehat\theta_{j+1}(0)\\
      \widehat q_{z,j+1}(0)
    \end{bmatrix}
    &=\mathbf I_j
    \begin{bmatrix}
      \widehat\theta_j(h_j)\\
      \widehat q_{z,j}(h_j)
    \end{bmatrix},\\[4pt]
    \mathbf I_j
    &=\begin{bmatrix}1&-R_j\\0&1\end{bmatrix}.
  \end{aligned}
  \label{eq:interface-matrix}
\end{equation}

Let $\boldsymbol\psi_j(z_j)\equiv
[\widehat\theta_j(z_j),\widehat q_{z,j}(z_j)]^{\mathsf T}$ denote the
temperature--flux state vector in layer $j$.  Then
\begin{equation}
  \begin{aligned}
    \boldsymbol\psi_N(h_N)
    &=\boldsymbol{\mathcal M}\boldsymbol\psi_1(0),\\
    \boldsymbol{\mathcal M}
    &=\mathbf L_N\mathbf I_{N-1}\mathbf L_{N-1}\cdots
      \mathbf I_1\mathbf L_1
      =\Phi\,\boldsymbol{\mathcal M}_0,\\
    \Phi&=\prod_{j=1}^{N}p_j
      =\exp\!\left(-\iu\sum_{j=1}^{N}
        \frac{\beta_jh_j}{k_{zz,j}}\right),\\
    \boldsymbol{\mathcal M}_0
    &=\mathbf L_{0,N}\mathbf I_{N-1}\mathbf L_{0,N-1}\cdots
      \mathbf I_1\mathbf L_{0,1}
      =\begin{bmatrix}m_{11}&m_{12}\\m_{21}&m_{22}\end{bmatrix}.
  \end{aligned}
  \label{eq:adiabatic-stack-factorization}
\end{equation}
All explicit shear dependence is contained in the common prefactor $\Phi$.
With an adiabatic rear surface, $\widehat q_{z,N}(h_N)=0$, and prescribed
heating flux $\widehat q_{\mathrm{top}}\equiv\widehat q_{z,1}(0)=\widehat q_p$,
\cref{eq:adiabatic-stack-factorization} gives
\begin{equation}
  \begin{gathered}
    0=\Phi\left(m_{21}\widehat\theta_{\mathrm{top}}
    +m_{22}\widehat q_{\mathrm{top}}\right),\\[3pt]
    \boxed{\;Z_{\mathrm{top}}\equiv
    \frac{\widehat\theta_{\mathrm{top}}}{\widehat q_{\mathrm{top}}}
    =-\frac{m_{22}}{m_{21}}\;}.
  \end{gathered}
  \label{eq:adiabatic-surface-impedance}
\end{equation}
The surface temperature $\widehat\theta_{\mathrm{top}}
=Z_{\mathrm{top}}\widehat q_{\mathrm{top}}$ contains no explicit shear phase.
This holds for arbitrary surface beam profiles and temporal waveforms under
linear Fourier diffusion, including continuous-wave FDTR and pulsed TDTR heating.
In layered samples, $k_{zz}$ can affect FDTR phase-frequency scans
and TDTR delay responses, including $-V_{\mathrm{in}}/V_{\mathrm{out}}$.
This sensitivity remains confined to the four invariants $(k_{zz},\Smat)$
for each otherwise unconstrained anisotropic layer.

For a non-adiabatic rear surface, convection and
radiation impose a Robin boundary condition.  Let the absolute
surface and ambient temperatures be
\begin{equation}
  T_{\mathrm{s}}=T_0+\theta_{\mathrm{s}},
  \qquad
  T_{\infty}=T_0+\theta_{\infty},
\end{equation}
where $T_0$ is the equilibrium temperature and the temperature rises are small
compared with $T_0$.  The heat flux leaving the rear surface is the sum of the
convective and radiative contributions,
\begin{equation}
  q_{\mathrm{rear}}
  =h_{\mathrm{c}}\left(T_{\mathrm{s}}-T_{\infty}\right)
  +\epsilon\sigma_{\mathrm{SB}}
   \left(T_{\mathrm{s}}^4-T_{\infty}^4\right).
  \label{eq:rear-convection-radiation}
\end{equation}
Linearization of the radiative term about $T_0$ gives
\begin{equation}
  T_{\mathrm{s}}^4-T_{\infty}^4
  \simeq
  4T_0^3\left(\theta_{\mathrm{s}}-\theta_{\infty}\right),
\end{equation}
and hence
\begin{equation}
  \begin{aligned}
    q_{\mathrm{rear}}
    &=H_{\mathrm{rear}}
      \left(\theta_{\mathrm{s}}-\theta_{\infty}\right),\\
    H_{\mathrm{rear}}
    &=h_{\mathrm{c}}+h_{\mathrm{r}},
    \qquad
    h_{\mathrm{r}}=4\epsilon\sigma_{\mathrm{SB}}T_0^3.
  \end{aligned}
  \label{eq:rear-effective-conductance}
\end{equation}
Here, $H_{\mathrm{rear}}$ is a scalar effective rear-surface heat-transfer
coefficient with units of $\mathrm{W\,m^{-2}K^{-1}}$. For a stationary ambient,
$\widehat\theta_{\infty}=0$.  Defining
$\widehat\theta_{\mathrm{rear}}\equiv\widehat\theta_N(h_N)$ and
$\widehat q_{\mathrm{rear}}\equiv\widehat q_{z,N}(h_N)$, the Fourier-domain boundary
condition reduces to
\begin{equation}
  \widehat q_{\mathrm{rear}}
  =H_{\mathrm{rear}}\widehat\theta_{\mathrm{rear}}.
  \label{eq:rear-robin-bc}
\end{equation}
Equivalently, the rear boundary condition becomes
\begin{align}
  0
  &=
  \begin{bmatrix}
    -H_{\mathrm{rear}}&1
  \end{bmatrix}
  \Phi
  \begin{bmatrix}
    m_{11}&m_{12}\\
    m_{21}&m_{22}
  \end{bmatrix}
  \begin{bmatrix}
    \widehat\theta_{\mathrm{top}}\\
    \widehat q_{\mathrm{top}}
  \end{bmatrix}
  \notag\\
  &=
  \Phi\left[
    \left(m_{21}-H_{\mathrm{rear}}m_{11}\right)
    \widehat\theta_{\mathrm{top}}
    +
    \left(m_{22}-H_{\mathrm{rear}}m_{12}\right)
    \widehat q_{\mathrm{top}}
  \right].
  \label{eq:rear-robin-expanded}
\end{align}
The front-surface
impedance is therefore
\begin{equation}
  \boxed{
  Z_{\mathrm{top}}
  \equiv
  \frac{\widehat\theta_{\mathrm{top}}}
       {\widehat q_{\mathrm{top}}}
  =
  \frac{m_{22}-H_{\mathrm{rear}}m_{12}}
       {H_{\mathrm{rear}}m_{11}-m_{21}}
  }.
  \label{eq:robin-surface-impedance}
\end{equation}
Thus, a spatially uniform convective--radiative rear boundary changes the
finite-thickness surface impedance but does not reintroduce explicit
dependence on the shear phase $\Phi$.

For a substrate whose other properties and coating are fixed, the exact
ambiguity follows from holding $(k_{zz},\Smat)$ fixed.  For any
coupling vector
$\bvec_{\mathrm{alt}}=(k_{xz,\mathrm{alt}},k_{yz,\mathrm{alt}})^{\mathsf T}$,
define
\begin{equation}
  \Kf(\bvec_{\mathrm{alt}})=
  \begin{bmatrix}
    \Smat+\bvec_{\mathrm{alt}}\bvec_{\mathrm{alt}}^{\mathsf T}/k_{zz}
      & \bvec_{\mathrm{alt}}\\
    \bvec_{\mathrm{alt}}^{\mathsf T} & k_{zz}
  \end{bmatrix}.
  \label{eq:equivalence-family}
\end{equation}
Every member shares $k_{zz}$ and $\Smat$, and hence the surface
response in \cref{eq:adiabatic-surface-impedance,eq:robin-surface-impedance}.
The corresponding Schur-equivalent conductivity matrix is
\begin{equation}
  \K_{\mathrm{eq}}=
  \begin{bmatrix}\Smat&\mathbf0\\\mathbf0^{\mathsf T}&k_{zz}\end{bmatrix}.
  \label{eq:zero-coupling-equivalent}
\end{equation}
\begin{figure}[t]
\centering
\begingroup
\renewcommand{\baselinestretch}{1}\selectfont
\resizebox{\textwidth}{!}{%
\begin{tikzpicture}[font=\footnotesize,>=Latex]

  \node[anchor=south west] at (-8.10,1.15)
    {\textbf{(a)}};
  \fill[blue!5] (-8.10,-2.20) rectangle (-3.55,0);
  \fill[yellow!35] (-8.10,-0.28) rectangle (-3.55,0);
  \draw[thick] (-8.10,0) -- (-3.55,0);
  \draw[gray] (-8.10,-0.28) -- (-3.55,-0.28);
  \node[anchor=west,font=\scriptsize,inner sep=1pt] at (-7.98,-0.14)
    {metal transducer};
  \draw[orange!80!black,very thick,domain=-0.75:0.75,samples=48]
    plot ({-5.70+\x},{0.12+0.46*exp(-4*\x*\x)});
  \foreach \x in {-6.15,-5.93,-5.70,-5.47,-5.25}
    \draw[orange!80!black,->] (\x,0.60) -- (\x,0.06);
  \node[orange!80!black,anchor=west] at (-4.85,0.42) {$q_p$};
  \node[anchor=south west] at (-8.06,0.08) {$z=0$};
  \draw[->,thick] (-7.92,-0.34) -- (-7.32,-0.34) node[anchor=west] {$x$};
  \draw[->,thick] (-7.92,-0.34) -- (-7.92,-0.94)
    node[anchor=north east,xshift=-2pt] {$+z$ (inward)};
  \node[draw,rounded corners,fill=white,inner sep=3pt] at (-5.35,-1.00)
    {$q_z=-k_{zz}\partial_z\theta-\bvec^{\mathsf T}\nabla_t\theta$};
  \node[anchor=center] at (-5.60,-1.80)
    {substrate: $\Kf$ or $\K_{\mathrm{eq}}$};

  \node[anchor=south west] at (-2.60,1.15)
    {\textbf{(b)}};
  \fill[yellow!35] (-2.20,-0.28) rectangle (2.30,0);
  \draw[thick] (-2.20,0) -- (2.30,0);
  \draw[gray] (-2.20,-0.28) -- (2.30,-0.28);
  \node[anchor=south west,inner sep=1pt] at (-2.18,0.08) {identical at $z=0$};
  \draw[->,thick] (-2.00,-0.35) -- (-2.00,-2.05) node[anchor=north] {$z$};
  \draw[gray!70,dashed,thick] (0.30,0) -- (0.30,-2.00);
  \draw[red!70!black,very thick] (0.30,0) -- (0.30,-0.28) -- (1.75,-2.00);
  \draw[gray!65,dashed] (0.30,-0.80) ellipse (0.36 and 0.085);
  \draw[gray!65,dashed] (0.30,-1.60) ellipse (0.36 and 0.085);
  \draw[red!70!black]   (0.74,-0.80) ellipse (0.36 and 0.085);
  \draw[red!70!black]   (1.41,-1.60) ellipse (0.36 and 0.085);
  \node[gray!75,anchor=east,inner sep=1pt] at (-0.12,-1.94)
    {$\K_{\mathrm{eq}}$};
  \node[red!70!black,anchor=west,inner sep=1pt] at (1.84,-1.94) {$\Kf$};
  \draw[<->,red!70!black] (0.30,-2.30) -- (1.75,-2.30);
  \node[red!70!black,anchor=north,inner sep=1pt] at (1.03,-2.50)
    {$(\Delta x,\Delta y)^{\mathsf T}=z_s\boldsymbol\gamma$};

  \node[anchor=south west] at (2.85,1.15)
    {\textbf{(c)}};
  \node[draw,rounded corners,fill=green!7,align=center,inner sep=6pt,
        anchor=north,text width=4.55cm] at (5.25,0.75)
  {$\displaystyle
    \Kf(\bvec_{\mathrm{alt}})=
    \begin{bmatrix}
      \Smat+\bvec_{\mathrm{alt}}\bvec_{\mathrm{alt}}^{\mathsf T}/k_{zz}
        &\bvec_{\mathrm{alt}}\\
      \bvec_{\mathrm{alt}}^{\mathsf T}&k_{zz}
    \end{bmatrix}$\\[5pt]
   any two-component column $\bvec_{\mathrm{alt}}$ at fixed
   $(k_{zz},\Smat)$\\[4pt]
   $\Longrightarrow$ identical single-orientation operator};

\end{tikzpicture}
}
\endgroup
\caption{Surface equivalence for a multi-layer sample.
(a) Heating and detection occur on the metal transducer, whose properties
and interface conductance are fixed. The substrate uses the complete
cross-plane heat flux. (b) For an isotropic transducer, the shear displacement
is zero throughout the coating and grows as $z_s\boldsymbol\gamma$ within
the substrate, where $\boldsymbol\gamma=\bvec/k_{zz}$ and $z_s$ is depth
below the transducer--substrate interface.
The substrate field is displaced relative to its Schur-equivalent field,
while the measured surface response is unchanged. (c) Every full anisotropic
thermal-conductivity tensor in the two-parameter family shares $k_{zz}$
and $\Smat$ and therefore gives the same surface response.}
\label{fig:geometry-equivalence}
\end{figure}
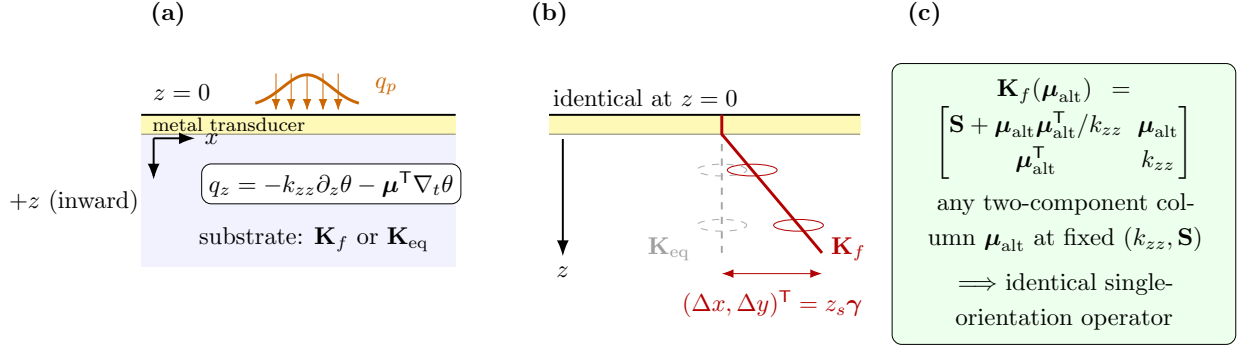

\subsection{Internal response}
\label{sec:multilayer-theorem-depth}

This cancellation applies to the front-surface response, but not to an
internal field at finite depth.  For $j\ge2$, the state at the top of layer
$j$ is
\begin{equation}
  \begin{aligned}
    \boldsymbol\psi_j(0)
    &=\Phi_{j-1}\mathbf I_{j-1}\mathbf L_{0,j-1}\cdots
      \mathbf I_1\mathbf L_{0,1}\boldsymbol\psi_1(0),\\
    \Phi_{j-1}&=\prod_{\ell=1}^{j-1}p_\ell .
  \end{aligned}
  \label{eq:internal-state-shear}
\end{equation}
We take the empty product to be $\Phi_0\equiv1$.
At a depth $z_j\in[0,h_j]$ within the same layer,
\begin{equation}
  \begin{aligned}
    \boldsymbol\psi_j(z_j)
    &=p_j(z_j)\mathbf L_{0,j}(z_j)\boldsymbol\psi_j(0)\\
    &=\Phi_j(z_j)\boldsymbol\psi_{\mathrm{eq},j}(z_j),\\
    p_j(z_j)&=\exp\!\left(-\frac{\iu\beta_jz_j}{k_{zz,j}}\right),
    \qquad
    \Phi_j(z_j)=\Phi_{j-1}p_j(z_j),
  \end{aligned}
  \label{eq:internal-depth-state}
\end{equation}
where $\boldsymbol\psi_{\mathrm{eq},j}$ is the state in the layerwise
Schur-equivalent stack.  Taking the temperature component gives
the explicit internal-temperature relation
\begin{equation}
  \boxed{
    \widehat\theta_j(\nuv,z_j;\omega)
    =\Phi_j(z_j)\widehat\theta_{\mathrm{eq},j}(\nuv,z_j;\omega)
  }.
  \label{eq:internal-temperature-shear}
\end{equation}
Thus $\Phi_1(0)=1$ at the measured surface, whereas an internal temperature
field retains the cumulative phase of every traversed layer and the partial
phase accumulated within layer $j$.

\subsection{Sensitivity-matrix SVD and local identifiability}
\label{sec:sensitivity-svd}

Sensitivity-matrix analysis tests which parameter combinations are
distinguishable near a specified tensor and experimental design
\cite{TuOng2023,ChenJiang2025}. For a positive signal $R$ and parameter
$\xi_j$, the conventional logarithmic sensitivity is
$S_{\xi_j}=\partial\ln R/\partial\ln\xi_j
=(\xi_j/R)\partial R/\partial\xi_j$.
Here we differentiate the selected signals directly, so phase and signed
or zero-valued off-diagonal entries require no logarithms.

The six parameters are
$\mathbf k=(k_{xx},k_{yy},k_{zz},k_{xy},k_{xz},k_{yz})^{\mathsf T}$.
Let $\mathbf y(\mathbf k)\in\mathbb R^{N_d}$ contain the selected
observations. For FDTR, these are the phases relative to the modulated
heating, stacked over frequencies, offset directions, and measured faces.
The physical sensitivity matrix is
\begin{equation}
  \Jmat
  =\frac{\partial\mathbf y}{\partial\mathbf k}
  =
  \begin{bmatrix}
    \dfrac{\partial y_1}{\partial k_{xx}}
      &\dfrac{\partial y_1}{\partial k_{yy}}
      &\cdots
      &\dfrac{\partial y_1}{\partial k_{yz}}\\[7pt]
    \dfrac{\partial y_2}{\partial k_{xx}}
      &\dfrac{\partial y_2}{\partial k_{yy}}
      &\cdots
      &\dfrac{\partial y_2}{\partial k_{yz}}\\
    \vdots&\vdots&\ddots&\vdots\\
    \dfrac{\partial y_{N_d}}{\partial k_{xx}}
      &\dfrac{\partial y_{N_d}}{\partial k_{yy}}
      &\cdots
      &\dfrac{\partial y_{N_d}}{\partial k_{yz}}
  \end{bmatrix}_{N_d\times6}.
  \label{eq:sensitivity-matrix}
\end{equation}
Each column varies one tensor component while the other five are fixed;
an off-diagonal variation changes both symmetric matrix positions.

We scale the six components using a fixed positive diagonal matrix
$\mathbf D=\diag(a_1,\ldots,a_6)$. Factoring the data covariance as
$\mathbf C_y=\mathbf L\mathbf L^{\mathsf T}$ gives the scaled, noise-weighted
matrix
\begin{equation}
  \widetilde{\Jmat}=\mathbf L^{-1}\Jmat\mathbf D.
  \label{eq:whitened-jacobian}
\end{equation}
For independent noise, row $m$ is divided by its standard deviation $s_m$.
In the example, diagonal entries use $a_{ii}=k_{ii,0}$ and
off-diagonal entries use $a_{ij}=\sqrt{k_{ii,0}k_{jj,0}}$, evaluated at the
nominal tensor. These fixed scales change singular values but preserve rank;
Supplementary Note~S2 gives the implementation.

The thin singular-value decomposition is
\begin{equation}
  \begin{aligned}
    \widetilde{\Jmat}
      &=\mathbf U\diag(\sigma_1,\ldots,\sigma_q)\mathbf V^{\mathsf T},\\
    q&=\min(N_d,6),
    \qquad \sigma_1\ge\cdots\ge\sigma_q\ge0.
  \end{aligned}
  \label{eq:sensitivity-svd}
\end{equation}
For a small tensor perturbation $\delta\mathbf k$, the linearized,
noise-weighted signal change is
\begin{equation}
  \delta\widetilde{\mathbf y}
  \equiv\mathbf L^{-1}\delta\mathbf y
  \simeq\sum_{j=1}^{q}
  \sigma_j\mathbf u_j
  \left(\mathbf v_j^{\mathsf T}\mathbf D^{-1}\delta\mathbf k\right).
  \label{eq:svd-perturbation}
\end{equation}
Thus $\mathbf v_j$ describes a combination of the six scaled tensor
changes; the corresponding physical direction is $\mathbf D\mathbf v_j$.
A zero singular value gives a null direction
$\Jmat\mathbf D\mathbf v_j=\mathbf0$, while a small nonzero value gives
a weakly sensed combination. Individual nonzero sensitivities therefore
do not establish independent identifiability.

The number of nonzero singular values is the local rank of the declared
forward map. We classify numerical nulls using
$\sigma_j/\sigma_1<10^{-8}$ together with derivative convergence and the
exact invariant family. This rank test is separate from practical
precision, which also depends on the noise and parameter scales.
For perturbations restricted to the retained scaled directions, the
linearized covariance in physical tensor entries is
\begin{equation}
  \mathbf C_{\mathbf k,\mathrm{id}}
  \simeq
  \mathbf D\mathbf V_r\diag(\sigma_1^{-2},\ldots,\sigma_r^{-2})
  \mathbf V_r^{\mathsf T}\mathbf D,
  \label{eq:svd-local-covariance}
\end{equation}
where $r$ is the retained rank and $\mathbf V_r$ contains its singular
vectors. This covariance is conditional on the other thermal inputs being
fixed and describes only the resolved directions. An exact null direction
remains undetermined and cannot be assigned a finite error bar from these
data alone.


\section{Results and discussion}
\label{sec:results}

\subsection{One orientation: Schur-equivalent FDTR and TDTR responses}
\label{sec:results-counterexample}

Consider a substrate with full anisotropic substrate tensor
\begin{equation}
  \K_0=
  \begin{bmatrix}
    10.4&2.0&-4.0\\
    2.0&7.0&4.0\\
    -4.0&4.0&8.6
  \end{bmatrix}
  \WmK.
  \label{eq:counterexample-tensor}
\end{equation}
With 80 nm Au transducer (detailed properties in Supplementary Note~S6).
The Schur complement is
$\Smat_0=\left[\begin{smallmatrix}
    8.5&3.9\\
    3.9&5.1
  \end{smallmatrix}\right]\,\WmK$.
Its Schur-equivalent conductivity matrix is
\begin{equation}
  \K_{\mathrm{eq},0}=
  \begin{bmatrix}
    8.5&3.9&0.0\\
    3.9&5.1&0.0\\
    0.0&0.0&8.6
  \end{bmatrix}
  \WmK.
  \label{eq:counterexample-equivalent}
\end{equation}
Thus the cross-plane coupling vector changes from
$(-4,4)^{\mathsf T}\,\WmK$ to zero while the in-plane block changes to
preserve $(k_{zz},\Smat_0)$.  Independent evaluations using the full
characteristic roots, complete cross-plane heat flux, and Schur formula agree
to numerical precision.

\begin{figure}[!htbp]
  \centering
  \linespread{1}\selectfont
  \includegraphics[width=0.95\textwidth]{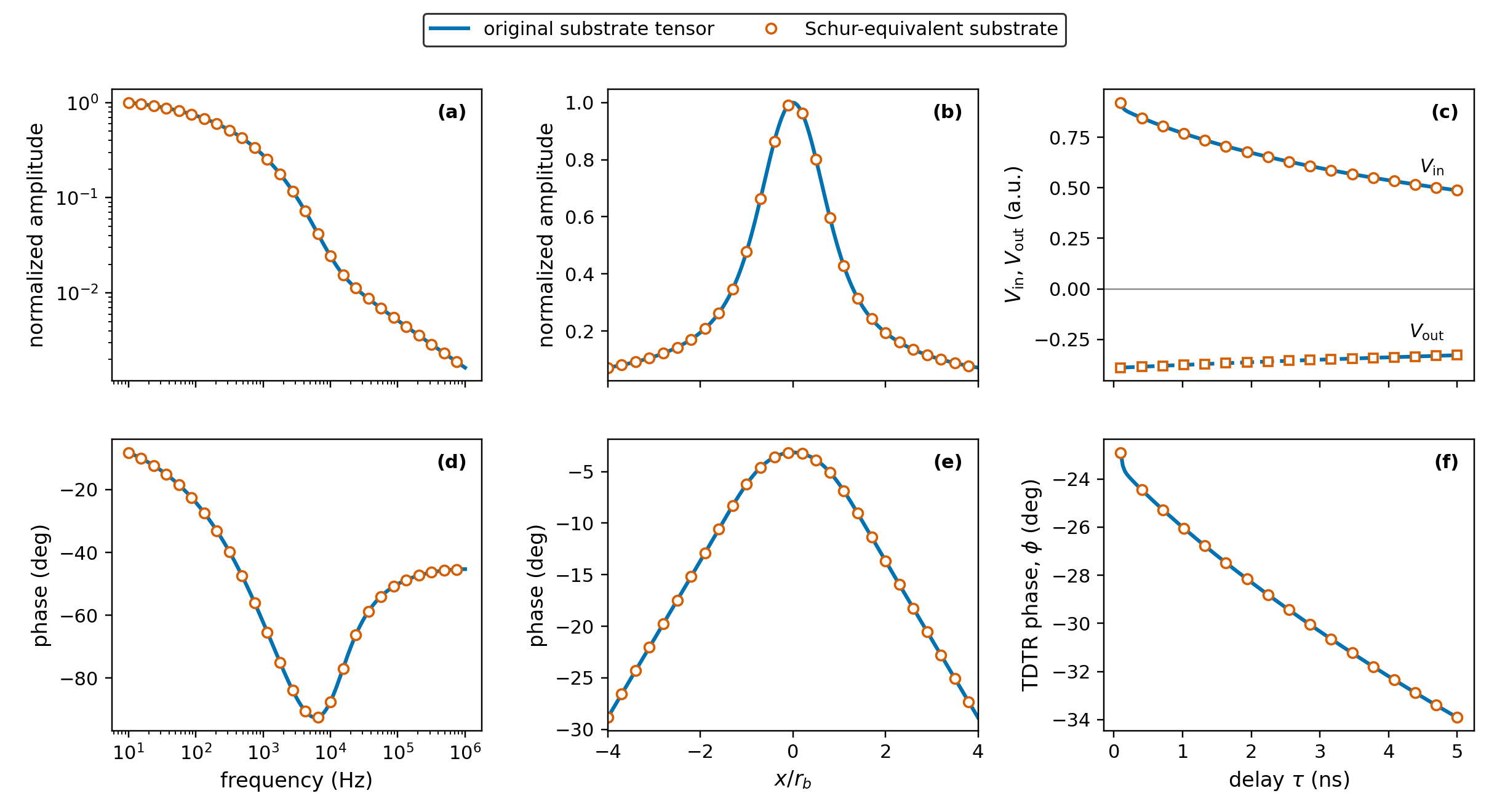}
  \caption{Single-orientation Schur equivalence for a coated substrate.
  Lines and open symbols compare $\K_0$ with its Schur-equivalent tensor
  for (a,d) FDTR frequency sweeps, (b,e) SDTR spatial scans, and
  (c,f) TDTR delay scans. Panels (a,b) show amplitude normalized by the
  original-tensor peak. Panel (c) shows $V_{\mathrm{in}}$ (solid/circles)
  and $V_{\mathrm{out}}$ (dashed/squares) in arbitrary units, scaled by the
  original complex-trace peak. Panels (d)--(f) show phase. Each tensor pair shares
  one normalization and overlaps to numerical precision. Stack and scan
  settings are given in Supplementary Note~S6.}
  \label{fig:equivalence-scans}
\end{figure}

\begin{figure}[!htbp]
  \centering
  \linespread{1}\selectfont
  \includegraphics[width=0.98\textwidth]{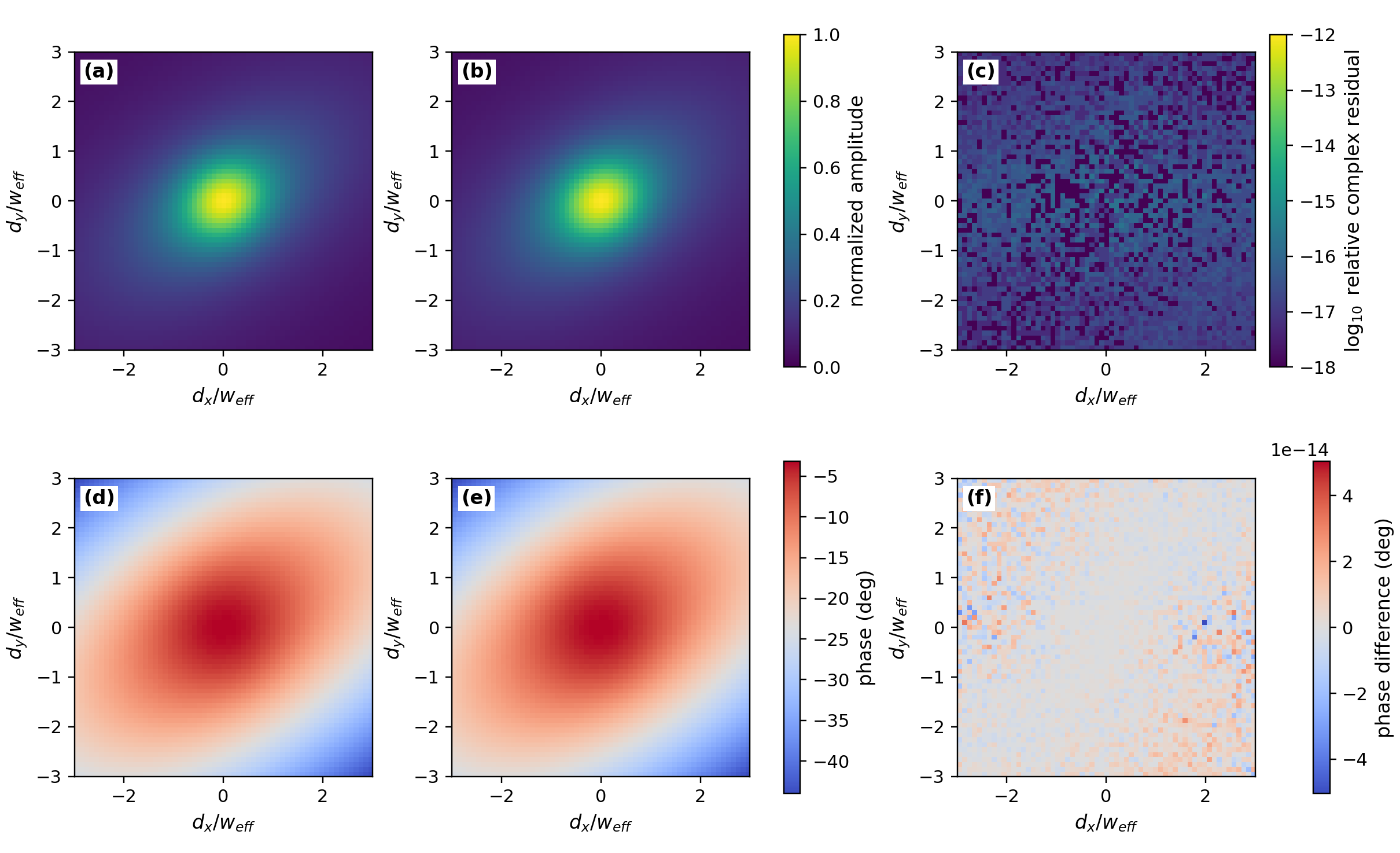}
  \caption{Complete two-dimensional transducer-surface response of the coated
  substrate with an adiabatic rear. Panels (a,b) show normalized amplitude
  and (d,e) show phase for the original and Schur-equivalent tensors,
  respectively. Panel (c) shows the logarithm of the relative complex
  residual, and (f) shows the phase residual. The amplitude maps share one
  normalization and the phase maps share one color scale. Residual panels show only
  floating-point and quadrature noise.  Phase residuals are wrapped and
  restricted to the reliable-signal mask defined in Supplementary Note~S6.}
  \label{fig:equivalence-maps}
\end{figure}

\Cref{fig:equivalence-scans} presents the comparison of the original tensor
and Schur-equivalent tensor for popular signal choices in FDTR and TDTR.
FDTR frequency sweeps and SDTR spatial scans~\cite{Jiang2022SpatialDomain}
therefore probe the same invariant set despite their different scan variables.  An
independent pulse-train TDTR reconstruction likewise gives indistinguishable
complex delay traces, quadratures, phase, and $-V_{\mathrm{in}}/V_{\mathrm{out}}$
ratios for the two tensors.  Supplementary Note~S3 separates the TDTR
equivalence residual from the harmonic-truncation error; Supplementary Note~S6
gives the remaining evaluation and convergence details.
\Cref{fig:equivalence-maps} extends the comparison to complete
two-dimensional surface maps used in SR-LIT tensor
analysis~\cite{Wang2024}. The amplitude and phase maps coincide for the two
tensors.

\FloatBarrier

\FloatBarrier
\subsection{Two orthogonal orientations: a common equivalent tensor}
\label{sec:two-face-equivalence}

We examine the information provided by two orthogonal surfaces
in two steps. First, we consider the full anisotropic tensor
$\K_0$ from \cref{eq:counterexample-tensor}. Second, we impose
the independently known constraint $k_{yz}=0$ and test recovery
of the remaining five components, including $k_{xz}$.
The input properties are specified in Supplementary Note~S6.

We label each face by its cross-plane axis. The $z$ and $y$
faces have registered local axes $(x,y;z)$ and $(z,x;y)$,
respectively, with the cross-plane axis listed last
(\cref{fig:three-face-geometry}). The same conductivity tensor
expressed in these two local frames is
\[
  \K^{(z)}=
  \begin{bmatrix}
    k_{xx}&k_{xy}&k_{xz}\\
    k_{xy}&k_{yy}&k_{yz}\\
    k_{xz}&k_{yz}&k_{zz}
  \end{bmatrix},
  \qquad
  \K^{(y)}=
  \begin{bmatrix}
    k_{zz}&k_{xz}&k_{yz}\\
    k_{xz}&k_{xx}&k_{xy}\\
    k_{yz}&k_{xy}&k_{yy}
  \end{bmatrix}.
\]
Their measured-face Schur complements are
\[
  \Smat_z=
  \begin{bmatrix}
    k_{xx}-\dfrac{k_{xz}^2}{k_{zz}} &
    k_{xy}-\dfrac{k_{xz}k_{yz}}{k_{zz}}\\[3pt]
    k_{xy}-\dfrac{k_{xz}k_{yz}}{k_{zz}} &
    k_{yy}-\dfrac{k_{yz}^2}{k_{zz}}
  \end{bmatrix},
  \qquad
  \Smat_y=
  \begin{bmatrix}
    k_{zz}-\dfrac{k_{yz}^2}{k_{yy}} &
    k_{xz}-\dfrac{k_{xy}k_{yz}}{k_{yy}}\\[3pt]
    k_{xz}-\dfrac{k_{xy}k_{yz}}{k_{yy}} &
    k_{xx}-\dfrac{k_{xy}^2}{k_{yy}}
  \end{bmatrix}.
\]
A common Schur-equivalent tensor $\K'$ must preserve both displayed
Schur matrices and their corresponding cross-plane conductivities:
$(k_{zz},\Smat_z)$ and $(k_{yy},\Smat_y)$. Thus
$k'_{zz}=k_{zz}$ and $k'_{yy}=k_{yy}$. Matching the lower-right
entry of $\Smat_z$ gives $k'_{yz}=\pm k_{yz}$. The unchanged
sign recovers $\K$; the opposite sign generally gives a distinct tensor.

To construct this second tensor, set $k'_{yz}=-k_{yz}$.
The off-diagonal Schur entries require
$k'_{xy}+(k_{yz}/k_{zz})k'_{xz}=S_{z,xy}$ and
$k'_{xz}+(k_{yz}/k_{yy})k'_{xy}=S_{y,zx}$.
Solving these two linear equations, then matching $S_{z,xx}$, gives
\[
\begin{aligned}
  k'_{xy}&=\frac{S_{z,xy}-(k_{yz}/k_{zz})S_{y,zx}}{\lambda},
  &k'_{xz}&=\frac{S_{y,zx}-(k_{yz}/k_{yy})S_{z,xy}}{\lambda},\\
  k'_{xx}&=S_{z,xx}+\frac{(k'_{xz})^2}{k_{zz}},
  &\lambda&=1-\frac{k_{yz}^2}{k_{yy}k_{zz}}>0.
\end{aligned}
\]
These entries also satisfy
$k'_{xx}-(k'_{xy})^2/k_{yy}=S_{y,xx}$, completing the equivalence
on both surfaces. The resulting common two-surface Schur-equivalent
conductivity tensor is
\begin{equation}
  \K_{\mathrm{eq}}^{(z,y)}=
  \begin{bmatrix}
    k'_{xx}&k'_{xy}&k'_{xz}\\
    k'_{xy}&k_{yy}&-k_{yz}\\
    k'_{xz}&-k_{yz}&k_{zz}
  \end{bmatrix}
  \simeq
  \begin{bmatrix}
    20.2&8.5&-10.0\\
    8.5&7.0&-4.0\\
    -10.0&-4.0&8.6
  \end{bmatrix}\,\WmK.
  \label{eq:two-face-equivalent-zy}
\end{equation}
The numerical matrix is obtained by substituting $\K_0$; its entries
are rounded to one decimal place. Both tensors are expressed in the same
global frame. Supplementary Note~S5 gives the corresponding branch
construction for the $z,x$ pair.

\begin{figure}[!htbp]
  \centering
  \linespread{1}\selectfont
  \includegraphics[width=0.98\textwidth]{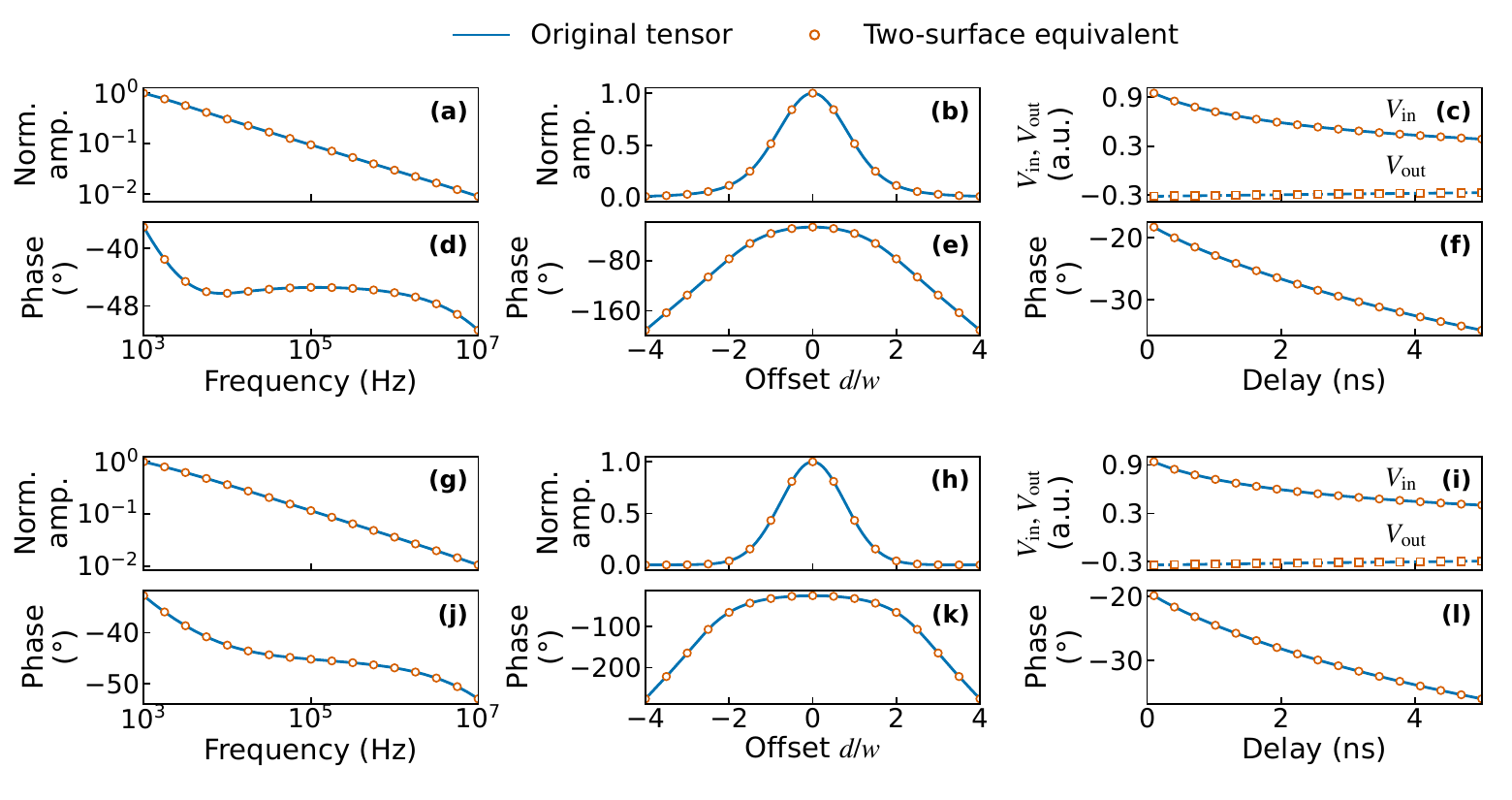}
  \caption{Common two-surface equivalence for $\K_0$ (lines) and
  $\K_{\mathrm{eq}}^{(z,y)}$ (symbols) on (a)--(f) the $z$ face and
  (g)--(l) the $y$ face. Columns compare FDTR frequency scans,
  SDTR spatial scans, and TDTR delay responses. Upper rows show
  normalized amplitude or $V_{\mathrm{in}}$ and $V_{\mathrm{out}}$ in
  arbitrary units; lower rows show phase. TDTR channels share the
  original complex-trace peak as their scale. Each tensor pair uses the
  same normalization. Measurement settings are
  given in Supplementary Note~S6.}
  \label{fig:two-surface-equivalence}

\end{figure}

\Cref{fig:two-surface-equivalence} compares independent full-tensor
calculations for $\K_0$ and $\K_{\mathrm{eq}}^{(z,y)}$. Their FDTR
frequency scans, SDTR spatial scans, and TDTR delay responses overlap
on both faces, confirming the common Schur equivalence. Varying the
signal choice therefore cannot remove this ambiguity.
Sufficiently informative two-face data can therefore have local rank six
without uniquely determining the tensor.

Two surfaces suffice if independent information fixes the coupling between
their cross-plane axes to zero. For the $z,y$ pair with known $k_{yz}=0$,
the five unknown components follow from
$k_{xy}=S_{z,xy}$, $k_{xz}=S_{y,zx}$, and
$k_{xx}=S_{z,xx}+k_{xz}^2/k_{zz}$, together with the measured
cross-plane conductivities $k_{zz}$ and $k_{yy}$. The two equivalent
branches then coincide. For the $z,x$ pair, the corresponding condition is
known $k_{xz}=0$, leaving $k_{yz}$ to be recovered. This zero-coupling
constraint requires independent material or orientation information;
registration of the surface axes alone does not impose it.

\begin{figure}[!htbp]
  \centering
  \linespread{1}\selectfont
  \includegraphics[width=0.85\textwidth]{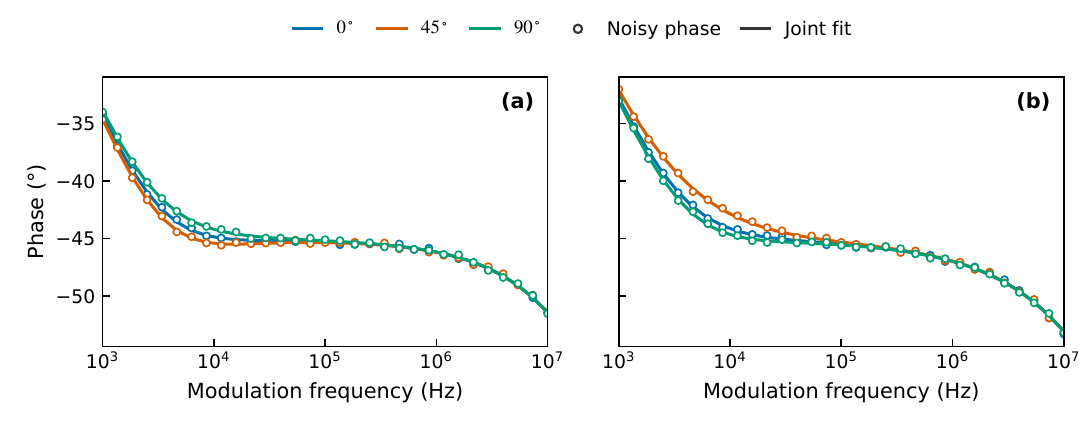}
  \caption{Five-component recovery from BO-FDTR phases on the (a) $z$ and
  (b) $y$ faces with $k_{yz}$ known to be zero. Symbols show noisy phase
  data and lines show the joint five-parameter fit. Colors identify the
  local offset azimuths. All other tensor entries are fitted simultaneously.}
  \label{fig:two-face-known-zero-recovery}
\end{figure}

To test this case, we set $k_{yz}=0$ in $\K_0$ and retain the other five
entries. Beam-offset FDTR (BO-FDTR)~\cite{Tang2021} phase scans at
$0^\circ$, $45^\circ$, and $90^\circ$ on each of the two faces use
31 frequencies from \qty{1}{\kilo\hertz} to \qty{10}{\mega\hertz}
and one-radius offsets. Adding independent relative phase noise bounded
by $\pm\qty{0.5}{\percent}$ gives 186 observations for a joint
five-parameter weighted nonlinear least-squares (NLLS) fit,
using inverse phase-noise variance weights~\cite{Tellinghuisen2009}
(\cref{fig:two-face-known-zero-recovery}). The fit
returns $\widehat\K=\left[\begin{smallmatrix}
10.6&2.1&-4.1\\2.1&7.0&0.0\\-4.1&0.0&8.6
\end{smallmatrix}\right]\,\WmK$, with a relative Frobenius error of
\qty{1.8}{\percent}.

\FloatBarrier


%
\newcommand{\tblleft}[2]{\parbox[t]{#1}{\raggedright\strut #2\strut}}
\newcommand{\tblcenter}[2]{\parbox[t]{#1}{\centering\strut #2\strut}}

\section{Guidance for determining the full anisotropic tensor}
\label{sec:limitations}
\label{sec:results-routes}
\label{sec:three-face-noisy-recovery}

\begin{figure}[!htbp]
  \centering
  \linespread{1}\selectfont
  \includegraphics[width=0.92\textwidth]{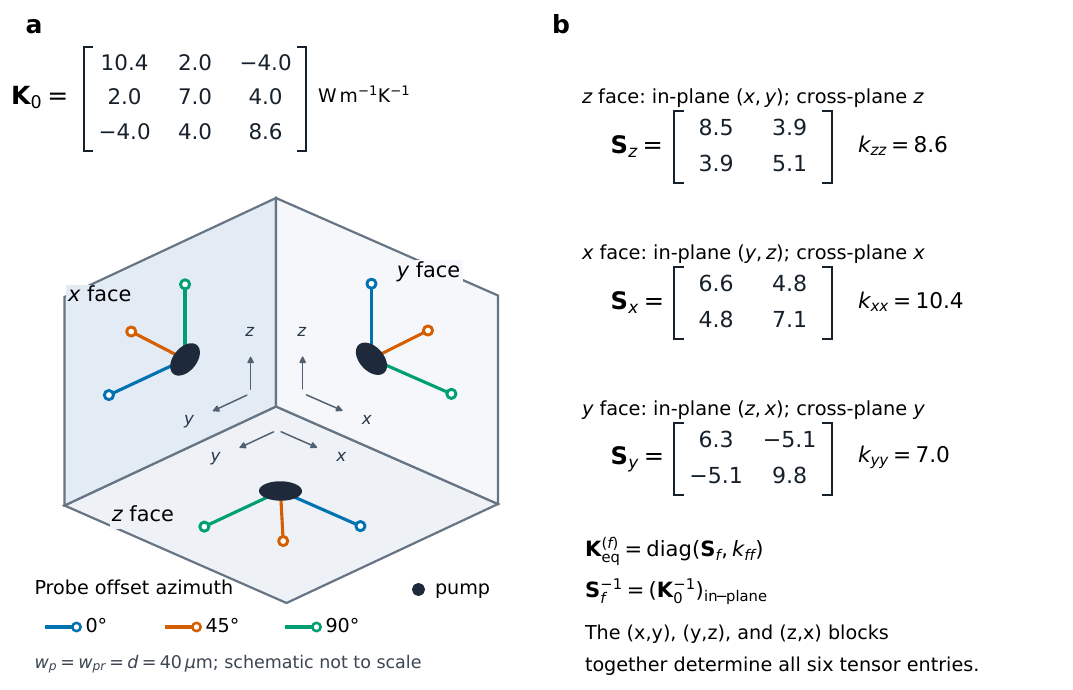}
  \caption{Three-face measurement arrangement for $\K_0$.
  (a) Registered face axes and probe offsets at $0^\circ$, $45^\circ$,
  and $90^\circ$ relative to the first in-plane axis. Each face carries
  the Au transducer, omitted from the schematic. Pump and probe radii
  are equal, and the offset is one radius. Scans are performed far from
  specimen edges. (b) Schur matrices and cross-plane conductivities in
  $\WmK$, rounded for display. Their inverse in-plane blocks jointly
  determine the global tensor.}
  \label{fig:three-face-geometry}
\end{figure}

In this section, we show that three mutually orthogonal
surfaces enable full-tensor recovery through simulated
BO-FDTR. Adding the third orthogonal surface removes the two-face ambiguity once
the three signed Schur matrices are resolved. Use face-local in-plane and
cross-plane axes $(x,y;z)$, $(y,z;x)$, and $(z,x;y)$ for the $z$, $x$,
and $y$ faces, respectively.
Each face has its own Schur-equivalent tensor
$\K_{\mathrm{eq}}^{(f)}=\diag(\Smat_f,k_{ff})$. The three sets of face
invariants constrain the same global tensor $\K$.
The block-inverse identity gives
\begin{equation}
  \Smat_f^{-1}=\Pmat_f^{\mathsf T}\K^{-1}\Pmat_f,
  \qquad f\in\{z,x,y\},
  \label{eq:three-face-resistivity-blocks}
\end{equation}
where the columns of $\Pmat_f$ are the two registered in-plane unit vectors.
Thus the three faces supply the $(x,y)$, $(y,z)$, and $(z,x)$ blocks of
$\K^{-1}$, which together determine all six entries. Inverting the assembled
thermal-resistivity matrix recovers $\K$. In particular, the additional $x$-face $(y,z)$
block distinguishes the two tensors constructed in
\cref{sec:two-face-equivalence}.
\Cref{fig:three-face-geometry} shows this arrangement for $\K_0$.
This establishes global uniqueness from the resolved face invariants;
the corresponding sensitivity ranks are summarized in
\cref{tab:orientation-rank}.

\begin{table}[!htbp]
\linespread{1}\selectfont
\caption{Sensitivity rank and tensor ambiguity for one, two, and three
measurement orientations.}
\label{tab:orientation-rank}
\footnotesize
\linespread{1.0}\selectfont
\setlength{\tabcolsep}{4pt}
\begin{tabular}{@{}lllll@{}}
\toprule
\tblleft{2.5cm}{Surface orientations}
  & \tblleft{3.6cm}{Signal choices}
  & \tblcenter{1.5cm}{Local rank}
  & \tblleft{4.4cm}{Recovered information}
  & \tblleft{3.0cm}{Remaining tensor ambiguity}\\
\midrule
\tblleft{2.5cm}{One orientation}
  & \tblleft{3.6cm}{FDTR: frequency, offset, and spot-size scans; TDTR: delay-time scans}
  & \tblcenter{1.5cm}{4}
  & \tblleft{4.4cm}{4: $k_{zz}$ and the three entries of the in-plane Schur complement $\Smat$}
  & \tblleft{3.0cm}{Continuous two-parameter family}\\
\tblleft{2.5cm}{Two orthogonal orientations}
  & \tblleft{3.6cm}{Same choices on both surfaces}
  & \tblcenter{1.5cm}{6}
  & \tblleft{4.4cm}{5: the remaining entry has two possible values}
  & \tblleft{3.0cm}{Two equivalent tensors in general}\\
\tblleft{2.5cm}{Three orthogonal orientations}
  & \tblleft{3.6cm}{Same choices on all three surfaces}
  & \tblcenter{1.5cm}{6}
  & \tblleft{4.4cm}{6: all six tensor entries}
  & \tblleft{3.0cm}{None; unique tensor recovery}\\
\bottomrule
\end{tabular}
\end{table}

For validation, we examine the same target full anisotropic tensor $\K_0$.
On each orthogonal face ($z$, $x$, and $y$),
the probe is displaced by one radius at local azimuths
$0^\circ$, $45^\circ$, and $90^\circ$
(\cref{fig:three-face-geometry}).
We concatenate the 31 phase points from three directional scans on each
of the three faces into a single vector containing
$3\times3\times31=279$ phase values.

To mimic experimental conditions, we add independent uniform noise
bounded by $\pm\qty{0.5}{\percent}$ of the noiseless phase magnitude.
We jointly fit all six tensor entries by weighted NLLS (inverse
phase-noise variance weights~\cite{Tellinghuisen2009}).
Other inputs are specified in Supplementary Note~S6.

To construct the sensitivity matrix, we use the 279 phase values
and six tensor parameters
$\mathbf k=(k_{xx},k_{yy},k_{zz},k_{xy},k_{xz},k_{yz})^{\mathsf T}$.
The resulting $279\times6$ matrix is evaluated at $\K_0$.
The SVD of its scaled, noise-weighted form,
$\widetilde{\mathbf J}=\mathbf U\boldsymbol\Sigma\mathbf V^{\mathsf T}$,
gives the normalized singular values
$(\sigma_1,\ldots,\sigma_6)/\sigma_1
=(1,\;0.570,\;0.518,\;0.425,\;0.309,\;0.186)$.
These six nonzero values establish local identifiability at $\K_0$.
The third face also removes the global ambiguity that can persist
with two orthogonal orientations, enabling unique recovery of the
full anisotropic tensor.

\begin{figure}[!htbp]
  \centering
  \linespread{1}\selectfont
  \includegraphics[width=\textwidth]{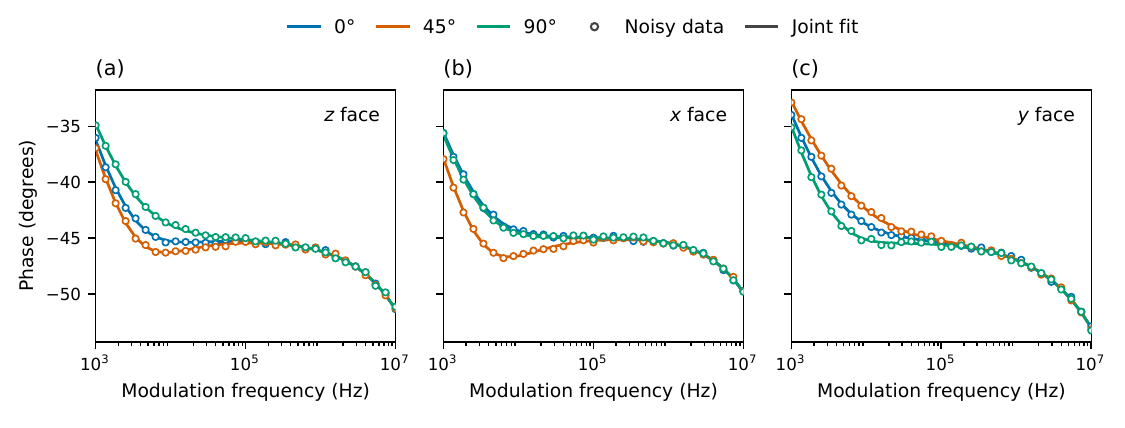}
  \caption{Joint six-entry recovery from noisy BO-FDTR phase-frequency scans on
  the (a) $z$, (b) $x$, and (c) $y$ faces. Open symbols are synthetic data
  with bounded relative phase noise; lines are the common NLLS fit.
  Colors denote the registered in-plane offset azimuths. All nine curves
  are fitted simultaneously using one global conductivity tensor.}
  \label{fig:three-face-recovery}
\end{figure}

The fit shown in \cref{fig:three-face-recovery} yields
$\widehat{\K}=\left[\begin{smallmatrix}
10.4&2.1&-3.9\\
2.1&7.0&4.0\\
-3.9&4.0&8.6
\end{smallmatrix}\right]\,\WmK$.
The largest relative entry error is \qty{6.6}{\percent} for
$k_{xy}$, while all other entries are within \qty{1.6}{\percent}
of their true values.
The larger relative error in $k_{xy}$ partly reflects its small
anisotropy coupling,
$|k_{xy}|/\sqrt{k_{xx}k_{yy}}
=2.0/\sqrt{10.4\times7.0}\simeq0.23$,
consistent with the higher fractional uncertainty reported for
weak off-diagonal couplings~\cite{Wang2025}.
To assess convergence stability, we tested eight initial guesses,
all of which converged to the same solution.

\FloatBarrier

\clearpage

\section{Conclusions}
\label{sec:conclusion}

We established an exact Schur-equivalent conductivity matrix for
single-orientation thermoreflectance measurements. The
temperature response depends on the substrate tensor only through $k_{zz}$
and the in-plane Schur complement
$\Smat=\A-\bvec\bvec^{\mathsf T}/k_{zz}$. Distinct tensors with the same
invariants therefore produce identical surface temperature responses. The
single-orientation upper limit is four independent tensor components.
We confirm this equivalence for both FDTR and TDTR data choices, including
phase, amplitude, and the TDTR ratio $-V_{\mathrm{in}}/V_{\mathrm{out}}$.
Varying measurement controls can improve precision but cannot remove the
exact nonuniqueness.
Two orthogonal surfaces can retain a common equivalent tensor. When the
coupling between their cross-plane axes is independently known to vanish,
they suffice to recover the remaining five components.
Finally, we show that three mutually orthogonal
orientations provide complementary information for full-tensor
recovery. Joint fitting of BO-FDTR data from these orientations
recovers all six tensor components, while the corresponding SVD
confirms their local identifiability.

\section*{Author declarations}

\subsection*{Conflict of interest}
The authors have no conflicts to disclose.

\clearpage
\setcounter{section}{0}
\setcounter{subsection}{0}
\setcounter{equation}{0}
\setcounter{figure}{0}
\setcounter{table}{0}
\renewcommand{\thesection}{S\arabic{section}}
\renewcommand{\thesubsection}{\thesection.\arabic{subsection}}
\renewcommand{\theequation}{S\arabic{equation}}
\renewcommand{\thefigure}{S\arabic{figure}}
\renewcommand{\thetable}{S\arabic{table}}
\renewcommand{\theHsection}{supplement.\arabic{section}}
\renewcommand{\theHsubsection}{supplement.\arabic{section}.\arabic{subsection}}
\renewcommand{\theHequation}{supplement.\arabic{equation}}
\renewcommand{\theHfigure}{supplement.\arabic{figure}}
\renewcommand{\theHtable}{supplement.\arabic{table}}

\begin{center}
  {\large\bfseries Supplementary Information for\par}
  \vspace{0.75em}
  {\large\itshape
  Identifiability of the three-dimensional anisotropic\\
  thermal-conductivity tensor in thermoreflectance measurements:\\
  Single-orientation limits and full-tensor recovery\par}
  \vspace{1em}
  {Dihui Wang and Heng Ban\par}
  {\itshape Department of Mechanical Engineering and Materials Science,
  University of Pittsburgh, Pittsburgh, PA, USA\par}
\end{center}


\section{Detailed Fourier reduction and multilayer impedance recursion}
\label{sec:supp-fourier}

This note gives the transform convention and intermediate algebra omitted
from the coated-stack derivation in the main text. The physical coordinates
are $(x,y,z)$, with $z$ positive into the transducer-coated specimen.  Only $x$ and $y$ are Fourier transformed; $z$ remains a
physical depth coordinate.  We use the spatial frequencies $u$ and $v$ in
cycles per unit length and collect them in the lateral spatial-frequency
vector
\begin{equation}
  \nuv\equiv\begin{bmatrix}u\\v\end{bmatrix},
  \qquad [\nuv]=\text{cycles per unit length}.
  \label{eq:supp-spatial-frequency-vector}
\end{equation}
Using the $e^{\iu\omega t}$ phasor
convention, let $\theta(x,y,z;\omega)$ be the complex harmonic temperature
amplitude, with $\omega=2\pi f$.  The forward and inverse
transforms are
\begin{subequations}
\label{eq:supp-fourier-pair}
\begin{align}
  \widehat\theta(\nuv,z;\omega)
  &=\int_{-\infty}^{\infty}\!\int_{-\infty}^{\infty}
    \theta(x,y,z;\omega)
    e^{-\iu 2\pi(ux+vy)}\,\dd x\,\dd y,
    \label{eq:supp-fourier-forward}\\
  \theta(x,y,z;\omega)
  &=\int_{-\infty}^{\infty}\!\int_{-\infty}^{\infty}
    \widehat\theta(\nuv,z;\omega)
    e^{\iu 2\pi(ux+vy)}\,\dd u\,\dd v.
    \label{eq:supp-fourier-inverse}
\end{align}
\end{subequations}
Here $\theta$ is the complex temperature amplitude, $f$ is the modulation
frequency in hertz, $\omega$ is angular frequency, and $\iu=\sqrt{-1}$.
The shorthand measure $\dd^2\nuv\equiv\dd u\,\dd v$ is used below.
The absence of a numerical prefactor in
\cref{eq:supp-fourier-inverse} follows from using cycles per unit length.  The
first derivative identity follows by integration by parts, assuming the
temperature field vanishes at lateral infinity:
\begin{align}
  \mathcal F_{x,y}\{\partial_x\theta\}
  &=\int_{-\infty}^{\infty}\!\int_{-\infty}^{\infty}
    \partial_x\theta\,e^{-\iu 2\pi(ux+vy)}\,\dd x\,\dd y\nonumber\\
  &=\int_{-\infty}^{\infty}
    \left[\theta e^{-\iu 2\pi(ux+vy)}\right]_{x=-\infty}^{x=\infty}\dd y
    +\iu 2\pi u\!\int_{-\infty}^{\infty}\!\int_{-\infty}^{\infty}
    \theta e^{-\iu 2\pi(ux+vy)}\,\dd x\,\dd y\nonumber\\
  &=\iu 2\pi u\,\widehat\theta.
  \label{eq:supp-integration-by-parts}
\end{align}
Repeated application gives
\begin{align}
  \partial_x&\longmapsto\iu 2\pi u,
  &\partial_y&\longmapsto\iu 2\pi v,
  &\partial_z&\longmapsto\partial_z,\nonumber\\
  \partial_{xx}&\longmapsto-4\pi^2u^2,
  &\partial_{yy}&\longmapsto-4\pi^2v^2,
  &\partial_{xy}&\longmapsto-4\pi^2uv,\nonumber\\
  \partial_{xz}&\longmapsto\iu 2\pi u\,\partial_z,
  &\partial_{yz}&\longmapsto\iu 2\pi v\,\partial_z,
  &\partial_{zz}&\longmapsto\partial_{zz}.
  \label{eq:supp-derivative-rules}
\end{align}

In coordinates parallel to the measured face, write the conductivity tensor
of each homogeneous layer as
\begin{equation}
  \Kf=
  \begin{bmatrix}
    \A&\bvec\\
    \bvec^{\mathsf T}&k_{zz}
  \end{bmatrix},
  \qquad
  \A=\begin{bmatrix}k_{xx}&k_{xy}\\k_{xy}&k_{yy}\end{bmatrix},
  \qquad
  \bvec=\begin{bmatrix}k_{xz}\\k_{yz}\end{bmatrix}.
  \label{eq:supp-tensor-partition}
\end{equation}
Here $\A\in\mathbb R^{2\times2}$ is the in-plane conductivity block,
$\bvec\in\mathbb R^{2\times1}$ is the cross-plane--in-plane coupling vector,
$\bvec^{\mathsf T}\in\mathbb R^{1\times2}$ is its transpose, and $k_{zz}$ is
the face-local cross-plane conductivity.
For spatially uniform properties, the harmonic heat equation
$\iu\omega C_v\theta=\nabla\!\cdot(\Kf\nabla\theta)$ expands to
\begin{equation}
  \iu\omega C_v\theta
  =\nabla_t^{\mathsf T}\A\nabla_t\theta
   +2\bvec^{\mathsf T}\nabla_t\partial_z\theta
   +k_{zz}\,\partial_z^2\theta,
  \qquad
  \nabla_t=\begin{bmatrix}\partial_x\\\partial_y\end{bmatrix}.
  \label{eq:supp-expanded-pde}
\end{equation}
Here $C_v$ is volumetric heat capacity and $\nabla_t$ is the in-plane
gradient operator.
Applying \cref{eq:supp-derivative-rules} term by term gives
\begin{align}
  \mathcal F_{x,y}\!\left\{
    \nabla_t^{\mathsf T}\A\nabla_t\theta\right\}
    &=-(2\pi)^2\nuv^{\mathsf T}\A\nuv\,\widehat\theta,
    \label{eq:supp-inplane-term}\\
  \mathcal F_{x,y}\!\left\{
    2\bvec^{\mathsf T}\nabla_t\partial_z\theta\right\}
    &=2\iu\left(2\pi\nuv^{\mathsf T}\bvec\right)\widehat\theta',
    \label{eq:supp-coupling-term}\\
  \mathcal F_{x,y}\!\left\{k_{zz}\,\partial_z^2\theta\right\}
    &=k_{zz}\,\widehat\theta''.
    \label{eq:supp-normal-term}
\end{align}
The prime notation denotes differentiation with respect to physical depth
$z$, with $(\nuv,\omega)$ held fixed:
\begin{equation*}
  \widehat\theta'\equiv\frac{\dd\widehat\theta}{\dd z},
  \qquad
  \widehat\theta''\equiv\frac{\dd^2\widehat\theta}{\dd z^2}.
\end{equation*}
Define
\begin{equation}
  \beta(\nuv)=2\pi\nuv^{\mathsf T}\bvec
       =2\pi(uk_{xz}+vk_{yz}),
  \label{eq:supp-beta}
\end{equation}
Both $\nuv$ and $\bvec$ are $2\times1$ columns, so $\beta$ is a scalar
coupling-weighted lateral Fourier coefficient.  The
transformed depth equation is
\begin{equation}
  k_{zz}\,\widehat\theta''+2\iu\beta\,\widehat\theta'
  -\left[(2\pi)^2\nuv^{\mathsf T}\A\nuv+\iu\omega C_v\right]
  \widehat\theta=0.
  \label{eq:supp-depth-ode}
\end{equation}
In component form, the in-plane quadratic term is
\begin{equation}
  \nuv^{\mathsf T}\A\nuv
  =k_{xx}u^2+2k_{xy}uv+k_{yy}v^2.
  \label{eq:supp-inplane-expanded}
\end{equation}

For each fixed $(\nuv,\omega)$, seek an exponential depth mode
$\widehat\theta(\nuv,z;\omega)=Ae^{sz}$, where $A$ is its complex amplitude and
$s$ is its complex propagation constant in inverse-length units.  Substitution
into \cref{eq:supp-depth-ode} gives
\begin{equation}
  k_{zz}s^2+2\iu\beta s
  -\left[(2\pi)^2\nuv^{\mathsf T}\A\nuv+\iu\omega C_v\right]=0.
  \label{eq:supp-characteristic}
\end{equation}
Equation \eqref{eq:supp-characteristic} is called the characteristic equation:
its roots are the values of $s$ that characterize the two independent
exponential solutions of the second-order depth equation.
Introduce the Schur complement
\begin{equation}
  \Smat=\A-\frac{\bvec\bvec^{\mathsf T}}{k_{zz}}.
  \label{eq:supp-schur}
\end{equation}
Writing $s=-\iu\beta/k_{zz}+\lambda$, where $\lambda$ is a shifted propagation
constant introduced only to complete the square, cancels the term linear in $\lambda$ and
reduces \cref{eq:supp-characteristic} to
\begin{equation}
  k_{zz}\lambda^2-\left[(2\pi)^2\nuv^{\mathsf T}\Smat\nuv
  +\iu\omega C_v\right]=0.
  \label{eq:supp-completed-square}
\end{equation}
Thus, the characteristic equation has the two roots
\begin{equation}
  s_{\pm}=-\frac{\iu\beta}{k_{zz}}\pm\sigma,
  \qquad
  \sigma=\sqrt{\frac{\iu\omega C_v+(2\pi)^2
  \nuv^{\mathsf T}\Smat\nuv}{k_{zz}}},
  \qquad \operatorname{Re}\sigma>0.
  \label{eq:supp-roots}
\end{equation}
The general solution of \cref{eq:supp-depth-ode} is therefore
\begin{equation}
  \widehat\theta(\nuv,z;\omega)
  =A_+(\nuv;\omega)e^{s_+z}
  +A_-(\nuv;\omega)e^{s_-z}.
  \label{eq:supp-general-depth-solution}
\end{equation}
Here $A_+$ and $A_-$ are the complex amplitudes fixed by the boundary and
interface conditions.  The two signs label the two independent depth modes.
The quantity $\sigma$ is the complex thermal decay constant, with its
square-root branch selected by $\operatorname{Re}\sigma>0$.  Consequently,
$s_+$ is the growing mode and $s_-$ is the decaying mode for increasing $z$.

The complete heat-flux amplitude is $\mathbf q=-\Kf\nabla\theta$.
Its cross-plane component relative to the measured face is
\begin{equation}
  q_z=-\bvec^{\mathsf T}\nabla_t\theta-k_{zz}\,\partial_z\theta,
  \label{eq:supp-normal-flux-real}
\end{equation}
which transforms to
\begin{equation}
  \widehat q_z=-\left(k_{zz}\frac{\dd}{\dd z}+\iu\beta\right)
  \widehat\theta.
  \label{eq:supp-normal-flux-transform}
\end{equation}
That is, $\widehat q_z\equiv\mathcal F_{x,y}\{q_z\}$.
For either depth root, define
$\widehat\theta_\pm=A_\pm e^{s_\pm z}$ and
$\widehat q_{z,\pm}=-(k_{zz}s_\pm+\iu\beta)\widehat\theta_\pm$.
The modal thermal admittance is the
flux-to-temperature ratio:
\begin{equation}
  \frac{\widehat q_{z,\pm}}{\widehat\theta_{\pm}}
  =-\left(k_{zz}s_{\pm}+\iu\beta\right)=\mp k_{zz}\sigma.
  \label{eq:supp-modal-admittance}
\end{equation}
The subscript $\pm$ associates each flux and temperature with the
corresponding propagation constant $s_\pm$.
The explicit term linear in $\bvec$ cancels only because
\cref{eq:supp-normal-flux-real} uses the complete cross-plane heat flux.

The sign convention takes $q_z>0$ into the coated specimen. For a
thermally thick substrate beneath the transducer, the decaying depth mode
sets the load impedance at the substrate top:
\begin{equation}
  Z_{\mathrm{sub}}^{\infty}
  =\frac{\widehat\theta_{\mathrm{sub}}(0)}
        {\widehat q_{z,\mathrm{sub}}(0)}
  =\frac{1}{k_{zz,\mathrm{sub}}\sigma_{\mathrm{sub}}}.
  \label{eq:supp-substrate-load}
\end{equation}
The coordinate in this equation starts at the transducer--substrate
interface. This load is combined with the interface resistance and
transducer propagation below to obtain the measured surface response.

\subsection{Adiabatic rear boundary and backward impedance recursion}

Consider $N\ge2$ finite, laterally homogeneous layers, numbered
$j=1,\ldots,N$ from the measured transducer surface to the substrate rear.  In layer $j$, reset the
local depth coordinate to $z_j=0$ at its top and let
$z_j\in[0,h_j]$, where $h_j$ is the layer thickness.  Denote its volumetric
heat capacity, cross-plane conductivity, coupling vector, and Schur complement by
$C_{v,j}$, $k_{zz,j}$, $\bvec_j$, and $\Smat_j$, respectively, and define
\begin{equation}
  a_j=\sigma_jh_j,
  \qquad
  Y_j=k_{zz,j}\sigma_j,
  \qquad
  p_j=e^{-\iu\beta_jh_j/k_{zz,j}},
  \label{eq:supp-layer-shorthand}
\end{equation}
where
\begin{equation}
  \sigma_j=\sqrt{\frac{\iu\omega C_{v,j}+(2\pi)^2
  \nuv^{\mathsf T}\Smat_j\nuv}{k_{zz,j}}},
  \qquad
  \beta_j=2\pi\nuv^{\mathsf T}\bvec_j,
  \qquad \operatorname{Re}\sigma_j>0.
  \label{eq:supp-layer-decay-beta}
\end{equation}
Also define the layerwise propagation constants
$s_{\pm,j}=-\iu\beta_j/k_{zz,j}\pm\sigma_j$.  Here $a_j$ is the
dimensionless thermal thickness, $Y_j$ is the characteristic thermal
admittance, and $p_j$ is the accumulated shear phase factor.
The state of temperature and cross-plane heat flux propagates through the layer as
\begin{equation}
  \boldsymbol\psi_j(h_j)
  =p_j\mathbf L_{0,j}\boldsymbol\psi_j(0),
  \qquad
  \mathbf L_{0,j}=
  \begin{bmatrix}
    \cosh a_j&-\sinh a_j/Y_j\\
    -Y_j\sinh a_j&\cosh a_j
  \end{bmatrix},
  \label{eq:supp-reduced-layer-matrix}
\end{equation}
where
$\boldsymbol\psi_j=(\widehat\theta_j,\widehat q_{z,j})^{\mathsf T}$.

For the last layer, the two-mode solution gives
\begin{equation}
  \widehat q_{z,N}(z)
  =-Y_NA_{+,N}e^{s_{+,N}z}
  +Y_NA_{-,N}e^{s_{-,N}z}.
  \label{eq:supp-last-layer-flux}
\end{equation}
The adiabatic rear condition $\widehat q_{z,N}(h_N)=0$ therefore requires
\begin{equation}
  A_{+,N}e^{s_{+,N}h_N}=A_{-,N}e^{s_{-,N}h_N},
  \qquad
  \frac{A_{+,N}}{A_{-,N}}=e^{-2\sigma_Nh_N}.
  \label{eq:supp-adiabatic-mode-relation}
\end{equation}
Thus a finite substrate with an adiabatic rear retains both modes;
$A_{+,N}=0$ gives the thermally thick substrate load in
\cref{eq:supp-substrate-load}.  At the top of the last layer,
\begin{equation}
  Z_N^{\mathrm{ad}}
  \equiv\frac{\widehat\theta_N(0)}{\widehat q_{z,N}(0)}
  =\frac{\coth a_N}{Y_N}
  =\frac{\coth(\sigma_Nh_N)}{k_{zz,N}\sigma_N}.
  \label{eq:supp-adiabatic-input-impedance}
\end{equation}
The coupling-weighted spatial frequency $\beta_N$ has canceled exactly.

The cancellation persists while the stack is built backward toward the
measured surface.  Define the input impedance at the top of any layer by
$Z_j=\widehat\theta_j(0)/\widehat q_{z,j}(0)$.  Across the scalar interface
resistance $R_j$, cross-plane flux is continuous,
$\widehat q_{z,j}(h_j)=\widehat q_{z,j+1}(0)\equiv\widehat q_{I,j}$, and
$\widehat\theta_j(h_j)=\widehat\theta_{j+1}(0)+R_j\widehat q_{I,j}$.
The load impedance seen by layer $j$ is therefore
\begin{equation}
  Z_{L,j}=R_j+Z_{j+1}.
  \label{eq:supp-interface-load-impedance}
\end{equation}
Substitution into \cref{eq:supp-reduced-layer-matrix} and solution for the
top impedance gives
\begin{equation}
  \boxed{
  Z_j=
  \frac{Z_{L,j}\cosh a_j+\sinh a_j/Y_j}
  {\cosh a_j+Z_{L,j}Y_j\sinh a_j}}.
  \label{eq:supp-impedance-recursion}
\end{equation}
The common factor $p_j$ cancels because an impedance is a
temperature-to-flux ratio.  Starting with the finite substrate load in
\cref{eq:supp-adiabatic-input-impedance}, or the thermally thick load in
\cref{eq:supp-substrate-load}, and applying
\cref{eq:supp-interface-load-impedance,eq:supp-impedance-recursion} for
$j=N-1,\ldots,1$ proves inductively that, with
$\widehat\theta_s\equiv\widehat\theta_1(0)$,
$\widehat q_p\equiv\widehat q_{z,1}(0)$, and $Z_s\equiv Z_1$,
\begin{equation}
  \widehat\theta_s(\nuv;\omega)
  =Z_1(\nuv;\omega)\widehat q_p(\nuv;\omega)
  \label{eq:supp-stack-surface-response}
\end{equation}
contains no explicit $\beta_j$.  Each anisotropic layer enters the
front-surface response through $(k_{zz,j},\Smat_j)$, not through $\bvec_j$
independently.

Finally, the finite adiabatic substrate load converges exponentially to
the thermally thick substrate load used in the same coated model.  Since
$Z_N^{\infty}=1/Y_N$,
\begin{equation}
  \left|\frac{Z_N^{\mathrm{ad}}}{Z_N^{\infty}}-1\right|
  =\left|\frac{2}{e^{2\sigma_Nh_N}-1}\right|
  \leq\frac{2}{e^{2\operatorname{Re}(\sigma_N)h_N}-1}.
  \label{eq:supp-adiabatic-error-bound}
\end{equation}
For a tolerance $\varepsilon$, it is sufficient that
\begin{equation}
  \operatorname{Re}(\sigma_N)h_N
  \geq\frac{1}{2}\ln\!\left(1+\frac{2}{\varepsilon}\right).
  \label{eq:supp-thermally-thick-criterion}
\end{equation}
The thresholds are $2.65$ for $1\%$ and $3.80$ for $0.1\%$.  For the
laterally uniform mode,
\begin{equation}
  \delta_N=\frac{1}{\operatorname{Re}\sigma_N}
  =\sqrt{\frac{k_{zz,N}}{\pi fC_{v,N}}},
  \label{eq:supp-normal-penetration-depth}
\end{equation}
where $\delta_N$ is the cross-plane thermal penetration depth of the laterally
uniform mode.  Thus the $1\%$ criterion is
$h_N/\delta_N\geq2.65$.  High frequency makes the
rear condition thermally irrelevant in the measured front response; it is
not required for the exact Schur cancellation under the adiabatic boundary.


\section{Sensitivity matrices for the six tensor components}
\label{sec:supp-numerical-coordinates}

The coated-sample FDTR calculations use the phase
$\phi_m=\arg H_m$ relative to the modulated heating, without subtracting
zero-offset phase. The six parameters are the physical tensor entries
$\mathbf k=(k_{xx},k_{yy},k_{zz},k_{xy},k_{xz},k_{yz})^{\mathsf T}$.
The sensitivity matrix is $J_{mj}=\partial\phi_m/\partial k_j$; each
row corresponds to a measured frequency and offset, and each column to
one tensor component. An off-diagonal derivative varies both symmetric
matrix positions together. All six columns are retained when testing
full-tensor identifiability.

For the Au-coated example, the matrix used for SVD is
\begin{equation}
  \widetilde J_{mj}=\frac{a_j}{s_m}J_{mj},\qquad
  a_{ii}=k_{ii,0},\qquad
  a_{ij}=\sqrt{k_{ii,0}k_{jj,0}}\quad(i\ne j),
  \label{eq:supp-physical-entry-scaling}
\end{equation}
where $s_m$ is the prescribed phase-noise standard deviation and the
subscript $0$ denotes the nominal tensor. The row weights account for the
noise at each observation; the fixed column scales permit dimensionless
comparison of the six sensitivity directions. Singular values are reported
relative to the largest, and values below $10^{-8}$ of the largest are
classified as numerical nulls. Scaling changes the singular values but
not the exact rank. Supplementary Note~S6 gives the data ordering,
noise model, and derivative checks.

The coated componentwise comparison instead uses the conventional
logarithmic sensitivity
$\partial\ln|\phi|/\partial\ln|k_{ij}|$ at the nonzero entries of
$\K_0$. Sign-preserving logarithmic perturbations at steps $10^{-3}$
and $5\times10^{-4}$ are combined by Richardson extrapolation after
continuous phase-branch alignment. These derivatives are with respect
to ordinary phase; center normalization is used only for the separately
specified amplitude channel. Supplementary Note~S4 fits a conditional
subset of three entries, using a scaled additive coordinate for the
entry whose nominal value can vanish. Its three-column matrix is
separate from the six-column full-tensor analysis.


\section{TDTR pulse-train extension and computational verification}
\label{sec:supp-tdtr}

\subsection{A TDTR delay trace is a sum of harmonic surface responses}

The single-orientation Schur equivalence is not restricted to sinusoidal
FDTR.  Under linear, time-invariant Fourier diffusion, the surface response
to an arbitrary temporal heat input is a superposition of the harmonic
surface responses derived in Supplementary Note~S1.  TDTR changes the
temporal pump waveform and additionally applies delayed probe sampling and
lock-in demodulation; all three operations are linear and independent of the
unknown conductivity tensor~\cite{Cahill2004,Schmidt2008}.

To make this statement explicit, let $T_r=1/f_r$ be the laser repetition
period, $\omega_r=2\pi f_r$ the repetition angular frequency,
$\omega_m=2\pi f_m$ the pump-modulation angular frequency, and $\tau$ the
pump--probe delay. Let $W_p$ and $W_d$ be the Fourier-domain pump and
detector weights, respectively. Define the unit-power harmonic surface
response
\begin{equation}
  H_{\K}(\omega,\mathbf d)
  =\int_{\mathbb R^2}W_d^*(\nuv)\,
  G_{s,\K}(\nuv;\omega)\,W_p(\nuv)\,
  e^{\iu2\pi\nuv^{\mathsf T}\mathbf d}\,\dd^2\nuv,
  \label{eq:supp-tdtr-harmonic-response}
\end{equation}
where $\mathbf d$ is the detector displacement and $G_{s,\K}$ is the
input impedance at the coated front surface for substrate tensor $\K$.
One complex lock-in sideband family of an amplitude-modulated pulse train has
the signed comb frequencies
\begin{equation}
  \omega_\ell=\omega_m+\ell\omega_r,
  \qquad \ell\in\mathbb Z.
  \label{eq:supp-tdtr-comb-frequencies}
\end{equation}
After delayed probe sampling and demodulation, its complex lock-in signal can
be written
\begin{equation}
  \mathcal V_{\K}(\tau)
  =\Gamma e^{\iu\omega_m\tau}
  \sum_{\ell=-\infty}^{\infty}
  c_\ell H_{\K}(\omega_\ell,\mathbf d)
  e^{\iu\ell\omega_r\tau}.
  \label{eq:supp-tdtr-comb-sum}
\end{equation}
Here $c_\ell$ contains the temporal pump-pulse spectrum, the finite probe-pulse
gate, and any other tensor-independent sideband weighting; $\Gamma$ contains
the thermoreflectance coefficient and detector gain.  A different lock-in
phase convention can absorb the common factor
$e^{\iu\omega_m\tau}$.  The real pump also has the conjugate modulation
sideband, and the unmodulated component is rejected by the lock-in.  Neither
detail changes the argument below because the Green-function equality holds
for every signed frequency.

For two tensors with the same $(k_{zz},\Smat)$, Supplementary Note~S1 gives
\begin{equation}
  G_{s,\K}(\nuv;\omega)
  =G_{s,\K_{\mathrm{eq}}}(\nuv;\omega)
  \quad\text{for every }(\nuv,\omega).
  \label{eq:supp-tdtr-termwise-green-equality}
\end{equation}
Consequently, the spatial integration in
\cref{eq:supp-tdtr-harmonic-response} and every term of
\cref{eq:supp-tdtr-comb-sum} are equal:
\begin{equation}
  \boxed{
  \mathcal V_{\K}(\tau)
  =\mathcal V_{\K_{\mathrm{eq}}}(\tau)
  \quad\text{for every delay }\tau.}
  \label{eq:supp-tdtr-equivalence}
\end{equation}
Thus $V_{\mathrm{in}}=\operatorname{Re}\mathcal V$,
$V_{\mathrm{out}}=\operatorname{Im}\mathcal V$, amplitude, phase, and the
usual ratio $-V_{\mathrm{in}}/V_{\mathrm{out}}$ are Schur-equivalent wherever
the ratio is defined. Delay scanning samples the same single-orientation
surface operator and cannot remove its two tensor null directions.

\subsection{Independent full-root computation}

We verified \cref{eq:supp-tdtr-equivalence} without evaluating both tensors
through a Schur-reduced kernel.  The original tensor and its Schur-equivalent
matrix were
\begin{equation}
  \K_0=
  \begin{bmatrix}
    10.4&2.0&-4.0\\
    2.0&7.0&4.0\\
    -4.0&4.0&8.6
  \end{bmatrix},
  \qquad
  \K_{\mathrm{eq},0}=
  \begin{bmatrix}
    8.5&3.9&0.0\\
    3.9&5.1&0.0\\
    0.0&0.0&8.6
  \end{bmatrix}
  \unit{\watt\per\meter\per\kelvin}.
  \label{eq:supp-tdtr-tensor-pair}
\end{equation}
For every signed comb frequency and lateral Fourier point, each tensor was
inserted separately into the complete characteristic equation
\begin{equation}
  k_{zz}s^2+2\iu\beta s-
  \left[(2\pi)^2\nuv^{\mathsf T}\A\nuv
  +\iu\omega C_v\right]=0,
  \qquad
  Y_{\mathrm{sub}}=-\left(k_{zz}s_-+\iu\beta\right),
  \label{eq:supp-tdtr-full-root-check}
\end{equation}
where the decaying root $s_-$ was selected and
$Y_{\mathrm{sub}}$ was calculated from the complete cross-plane heat flux.  This
directly retains $\A$ and $\bvec$ in the numerical calculation;
the cancellation is therefore a result of the solution, not an input
assumption.

The finite adiabatic substrate contributes the input impedance
$Z_{\mathrm{sub}}=[Y_{\mathrm{sub}}\tanh(\sigma h_{\mathrm{sub}})]^{-1}$,
where $\sigma=-s_- -\iu\beta/k_{zz}$. The interface resistance is added
to this load before propagation through the transducer.

The material stack was the same as the finite-stack comparison in the main
text: a \qty{100}{\nano\meter} isotropic transducer with
$k=\qty{100}{\watt\per\meter\per\kelvin}$ and
$C_v=\qty{2.4e6}{\joule\per\meter\cubed\per\kelvin}$, a scalar interface
resistance of \qty{2e-8}{\meter\squared\kelvin\per\watt}, and a
\qty{1.8}{\milli\meter} substrate with
$C_v=\qty{1.95e6}{\joule\per\meter\cubed\per\kelvin}$ and an adiabatic rear.
Equal Gaussian pump and probe radii of
$40/\sqrt2\,\unit{\micro\meter}$ preserve the
$\qty{40}{\micro\meter}$ effective spatial weighting of the FDTR comparison,
so only the temporal source and TDTR sampling are changed.  We used
$f_r=\qty{80}{\mega\hertz}$, $f_m=\qty{10}{\mega\hertz}$, Gaussian pump and
probe temporal full widths at half maximum of \qty{2}{\pico\second}, and 161
delays from \qty{0.10}{\nano\second} to \qty{5.0}{\nano\second}.  The lateral
integral used $24\times24$ Gauss--Hermite nodes, and the final reconstruction
retained $\ell=-16384,\ldots,16384$.

The maximum comb-tooth discrepancy, scaled by the largest harmonic response,
was below $1\times10^{-16}$.  After pulse accumulation, the maximum complex
delay-trace discrepancy was $3.1\times10^{-16}$, the maximum reliable-signal
phase discrepancy was $1.6\times10^{-14}$ degrees, and the peak-scaled TDTR
ratio discrepancy was $3.8\times10^{-16}$.  Doubling the harmonic cutoff from
8192 to 16384 changed the original-tensor trace by $5.9\times10^{-8}$; this
comb-truncation convergence is reported separately from the much smaller
Schur-equivalence residual.

\begin{figure}[t]
  \centering
  \includegraphics[width=\linewidth]{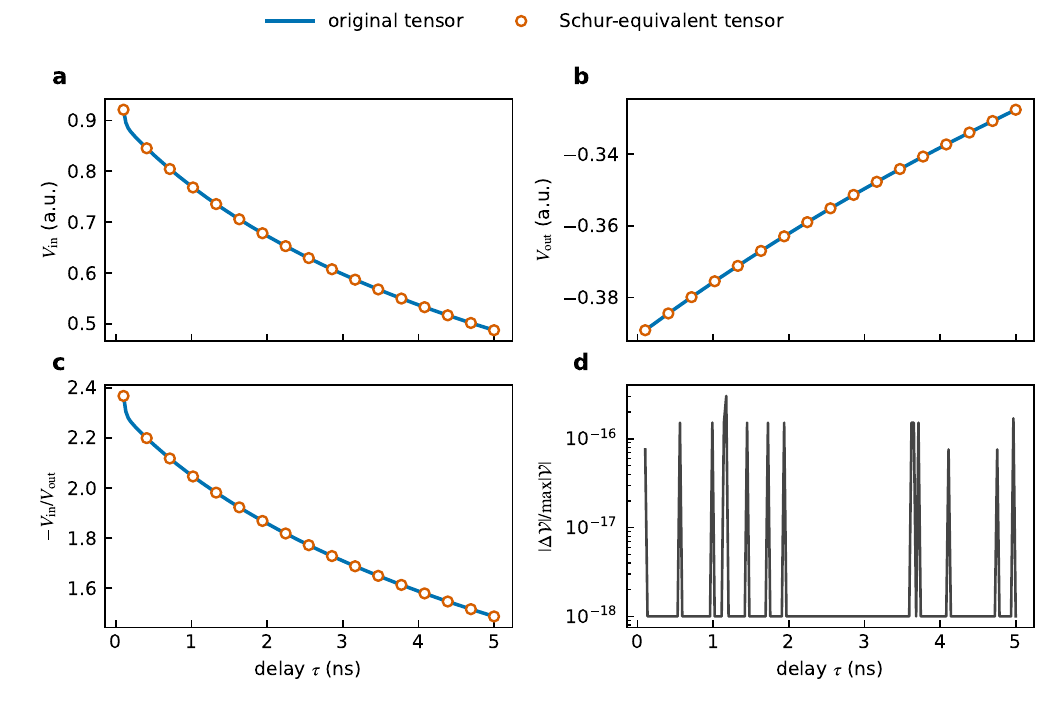}
  \caption{Independent pulse-train TDTR verification of the
  Schur-equivalent tensor pair in \cref{eq:supp-tdtr-tensor-pair}, using
  \qty{80}{\mega\hertz} repetition and \qty{10}{\mega\hertz} modulation. Panels
  (a) and (b) show $V_{\mathrm{in}}$ and $V_{\mathrm{out}}$ in arbitrary
  units, scaled by the common peak complex amplitude.  Panel (c) shows
  $-V_{\mathrm{in}}/V_{\mathrm{out}}$ where
  $|V_{\mathrm{out}}|$ exceeds 8\% of its peak.  Panel (d) shows the complex
  signal difference.  The original tensor is the solid line and the
  equivalent tensor is shown by sparse open circles.  Their complete delay
  traces agree at roundoff precision.}
  \label{fig:supp-tdtr-verification}
\end{figure}

The extension assumes linear small-signal thermoreflectance, local Fourier
diffusion with constant properties, surface heating, surface detection on the
same planar orientation, laterally homogeneous layers, the complete cross-plane heat
flux, and scalar interface resistances.  Volumetric absorption inside
the unknown anisotropic layer, a buried or rear detector, an anisotropic or
patterned interface, quasiballistic transport, or temperature-dependent
properties can introduce depth information or nonlinear frequency mixing and
must be analyzed separately.

\FloatBarrier

\section{Conditional SVD and propagated parameter uncertainty}
\label{sec:supp-conditional-uncertainty}

This coated-sample control fits only three tensor components while
holding the other three fixed. Its three-column sensitivity matrix
therefore addresses conditional recovery, not the six-unknown
full-tensor problem. The coating, interface, and scan settings are those
of the componentwise comparison in Supplementary Note~S6.

This note separates three quantities that use similar notation in parts of
the thermoreflectance literature.  We reserve $\sigma_i$ for the singular
values of a sensitivity matrix and $\sigma_i/\sigma_1$ for their normalized
ratios.  The propagated one-standard-deviation uncertainty of a fitted
conductivity component is instead denoted by
\begin{equation}
  u(k_j)=\sqrt{(\mathbf C_k)_{jj}}.
  \label{eq:supp-standard-uncertainty}
\end{equation}
Thus $u(k_j)$ has units of thermal conductivity and is neither a singular
value nor a normalized singular value.

\subsection{Conditional coordinates and error propagation}

The paired calculation uses
\begin{equation}
  \K(e)=
  \begin{bmatrix}
    30.0&6.8&27.0\\
    6.8&9.2&e\\
    27.0&e&30.0
  \end{bmatrix}
  \unit{\watt\per\meter\per\kelvin},
  \qquad e\in\{5,0\}.
  \label{eq:supp-conditional-tensors}
\end{equation}
Only $(k_{yy},k_{zz},k_{yz})$ are fitted.  The entries
$(k_{xx},k_{xy},k_{xz})=(30,6.8,27)\,
\unit{\watt\per\meter\per\kelvin}$ are controlled inputs.  Both tensors use
the dimensionless coordinates
\begin{equation}
  q_{yy}=\ln\frac{k_{yy}}{k_{yy,0}},\qquad
  q_{zz}=\ln\frac{k_{zz}}{k_{zz,0}},\qquad
  q_{yz}=\frac{k_{yz}-k_{yz,0}}
  {\sqrt{k_{yy,0}k_{zz,0}}}.
  \label{eq:supp-conditional-coordinates}
\end{equation}
The scaled-additive $q_{yz}$ remains locally defined when $k_{yz,0}=0$.

Let $\mathbf g(\mathbf q,\mathbf p)$ be the stacked vector of 81 spatial-scan
and 81 frequency-scan phase values in degrees, where $\mathbf p$ contains the
controlled inputs. Both scans use the unwrapped phase of the complex
response relative to the heating, without zero-offset phase subtraction.  Define the dimensional phase Jacobians
\begin{equation}
  \mathbf J_q=\frac{\partial\mathbf g}{\partial\mathbf q},
  \qquad
  \mathbf J_p=\frac{\partial\mathbf g}{\partial\mathbf p}.
  \label{eq:supp-conditional-jacobians}
\end{equation}
The SVD calculation stores
$\mathbf J_q^{\log}=\partial\ln|\boldsymbol\phi|/\partial\mathbf q$.
At the nonzero baseline phase, the chain rule converts it to the raw-phase
Jacobian used for uncertainty propagation,
\begin{equation}
  \mathbf J_q=\operatorname{diag}(\boldsymbol\phi_{\rm deg})
  \mathbf J_q^{\log},
  \label{eq:supp-log-to-raw-phase}
\end{equation}
with the same conversion applied to each controlled-input column.

For the unweighted least-squares estimator of Yang, Ziade, and
Schmidt,\cite{Yang2016Uncertainty} the first-order covariance is
\begin{align}
  \mathbf H&=(\mathbf J_q^{\mathsf T}\mathbf J_q)^{-1}
  \mathbf J_q^{\mathsf T},
  \label{eq:supp-ols-estimator}\\
  \mathbf C_{\rm eff}
  &=\mathbf C_y+\mathbf J_p\mathbf C_p\mathbf J_p^{\mathsf T},
  \label{eq:supp-effective-covariance}\\
  \mathbf C_q&=\mathbf H\mathbf C_{\rm eff}\mathbf H^{\mathsf T}.
  \label{eq:supp-jia-yang-covariance}
\end{align}
Here $\mathbf C_y$ and $\mathbf C_p$ are the measurement and controlled-input
covariances.  The plus sign in \cref{eq:supp-effective-covariance} follows
because independent measurement and controlled-input errors both add
data-space variance.  The physical covariance is
\begin{equation}
  \mathbf C_k=\mathbf D\mathbf C_q\mathbf D^{\mathsf T},\qquad
  \mathbf D=\operatorname{diag}
  \left(k_{yy,0},k_{zz,0},\sqrt{k_{yy,0}k_{zz,0}}\right).
  \label{eq:supp-physical-covariance}
\end{equation}

\subsection{Matched uncertainty assumptions}

The comparison assigns the same independent one-standard-deviation
uncertainties to both tensors: $\qty{0.1}{\degree}$ at every phase point;
\qty{5}{\percent} for the three controlled tensor entries, the two beam
radii, both layer thicknesses, and spatial-position calibration;
\qty{20}{\percent} for transducer conductivity and interface conductance;
\qty{2}{\percent} for transducer and substrate heat capacities; and
\qty{0.1}{\percent} for frequency calibration.  The percentage priors are
implemented in sign-preserving logarithmic coordinates.  These are local,
independent Gaussian assumptions used for a controlled comparison, not a
universal instrument specification.

\subsection{Results and interpretation}

\Cref{tab:supp-conditional-uncertainty} reports three scenarios.  The
measurement-only case enforces all controlled inputs exactly.  The second
propagates experimental controls while retaining the three complementary
tensor entries as exact.  The third also assigns those nominally fixed tensor
entries their declared \qty{5}{\percent} uncertainty.

\begin{table}[htbp]
\centering
\caption{Propagated one-standard-deviation parameter uncertainties.  Values
are in \unit{\watt\per\meter\per\kelvin}; parentheses give uncertainty
relative to the nonzero nominal component.  A relative percentage is
undefined for nominal $k_{yz}=0$.}
\label{tab:supp-conditional-uncertainty}
\small
\resizebox{\textwidth}{!}{%
\begin{tabular}{@{}l@{\hspace{0.8em}}lccc@{}}
\toprule
Scenario & $k_{yz}$ & $u(k_{yy})$ & $u(k_{zz})$ & $u(k_{yz})$\\
\midrule
Measurement only
  & $5$ & $0.0218\ (0.237\%)$ & $0.0104\ (0.0347\%)$ & $0.00783\ (0.157\%)$\\
Measurement only
  & $0$ & $0.0179\ (0.194\%)$ & $0.0191\ (0.0635\%)$ & $0.00804$\\
Experiment controls; tensor entries exact
  & $5$ & $6.47\ (70.3\%)$ & $1.73\ (5.77\%)$ & $2.81\ (56.2\%)$\\
Experiment controls; tensor entries exact
  & $0$ & $5.14\ (55.8\%)$ & $6.86\ (22.9\%)$ & $2.32$\\
All declared controls
  & $5$ & $6.47\ (70.3\%)$ & $3.91\ (13.0\%)$ & $2.86\ (57.2\%)$\\
All declared controls
  & $0$ & $5.14\ (55.9\%)$ & $7.71\ (25.7\%)$ & $2.35$\\
\bottomrule
\end{tabular}
}
\end{table}

For nominal $k_{yz}=0$, the final $u(k_{yz})=2.35\,
\unit{\watt\per\meter\per\kelvin}$ is \qty{14.2}{\percent} of the natural
coordinate scale $\sqrt{k_{yy,0}k_{zz,0}}$.  Relative to the nonzero case,
the all-control uncertainties change by factors $(0.794,1.973,0.822)$ for
$(k_{yy},k_{zz},k_{yz})$.  The zero case therefore approximately doubles the
$k_{zz}$ uncertainty, but it is not uniformly less precise in every reported
component.

The unweighted logarithmic-phase singular values are
$(18.05,4.769,1.036)$ for $k_{yz}=5$ and $(159.6,2.638,1.420)$ for
$k_{yz}=0$.  Although every raw value exceeds $0.1$, their normalized spectra
are respectively
\begin{equation}
  (1,0.264,0.0574),\qquad (1,0.0165,0.00889).
  \label{eq:supp-relative-spectra}
\end{equation}
The leading modes contain \qty{93.2}{\percent} and \qty{99.965}{\percent} of
the squared raw sensitivity.  Their squared loadings in
$(q_{yy},q_{zz},q_{yz})$ are $(0.006,0.945,0.049)$ and
$(0.022,0.548,0.429)$, respectively.  Consequently, the large value $159.6$
is not a singular value belonging to $k_{yy}$; it describes a mixed direction
dominated by $q_{zz}$ and $q_{yz}$.

A raw absolute cutoff such as $\sigma_i>0.1$ is not portable because singular
values change with data count, observable definition, and parameter scaling.
Normalized singular values expose mode imbalance but do not themselves define
acceptable parameter precision.  Indeed, a relative cutoff of $0.1$ would
discard sensitivity directions in these three-parameter matrices even though both
measurement-only conditional fits in
\cref{tab:supp-conditional-uncertainty} have finite uncertainty.

For a covariance-aware diagnostic, let
$\mathbf C_{\rm eff}=\mathbf L\mathbf L^{\mathsf T}$ and form the whitened
Jacobian $\widetilde{\mathbf J}_q=\mathbf L^{-1}\mathbf J_q$.  With all
declared controls, its relative singular spectra are
$(1,0.437,0.286)$ and $(1,0.322,0.259)$, with condition numbers $3.50$ and
$3.87$.  A fit that actually minimizes the corresponding covariance-weighted
residual has covariance
$(\mathbf J_q^{\mathsf T}\mathbf C_{\rm eff}^{-1}\mathbf J_q)^{-1}$; this is
a different estimator from the unweighted Yang--Ziade--Schmidt propagation in
\cref{eq:supp-jia-yang-covariance}.  Practical measurability is therefore best
reported using the declared covariance, propagated $u(k_j)$, and an explicit
acceptable-precision criterion.  Structural rank remains a separate analytic
or tight-relative-tolerance question.


\section{Global ambiguities that a local SVD cannot detect}
\label{sec:supp-global-svd}

\subsection{Centered circular data determine an unordered in-plane pair}

Consider the diagonal control
\begin{equation}
  \K=\operatorname{diag}(k_x,k_y,k_z)
\end{equation}
with an isotropic transducer and co-centered circular pump and detector
on the $z$ face.  At zero
offset, the Fourier-domain signal can be written schematically as
\begin{equation}
  H(k_x,k_y;\omega)
  =\iint_{\mathbb R^2}
  W(u^2+v^2)\,
  \mathcal G(u^2+v^2,k_xu^2+k_yv^2;\omega)\,\dd u\,\dd v,
  \label{eq:supp-circular-signal}
\end{equation}
where the circular pump and detector factors are collected in the radial
weight $W$, and $\mathcal G$ also contains the fixed cross-plane and layer
properties.  The substitution $(u,v)\mapsto(v,u)$ gives the exact exchange
symmetry
\begin{equation}
  H(k_x,k_y;\omega)=H(k_y,k_x;\omega)
  \qquad\text{for every }\omega.
  \label{eq:supp-inplane-exchange}
\end{equation}
Consequently, centered circular frequency data can determine the two in-plane
eigenvalues only as an unordered pair; they do not assign the larger value to
the laboratory $x$ or $y$ axis.

This discrete global ambiguity is compatible with a full-rank local
Jacobian. The coated control uses the \qty{100}{\nano\meter} transducer
and interface of Supplementary Note~S6, an effective circular radius of
\qty{40}{\micro\meter}, and 101 frequencies from \qty{10}{\hertz} to
\qty{1}{\mega\hertz}. At
$(k_x,k_y,k_z)=(50,10,10)\,\WmK$, only $k_x$ and $k_y$ are varied;
$k_z$ and all off-diagonal entries are fixed. The unweighted
$101\times2$ matrix
$\partial\ln|\phi(f)|/\partial(\ln k_x,\ln k_y)$ has singular values
$(2.4325,0.3849)$. These are the raw stacked values, without a division
by the square root of the number of observations. Exchanging $k_x$
and $k_y$ changes the complex trace by less than
$2\times10^{-16}$ relative to its peak. Thus each branch is locally
regular, although the two parameter vectors produce the same coated
FDTR phase response.

Baseline dependence follows from the same symmetry.  Define the
antisymmetric logarithmic path
\begin{equation}
  k_x(\delta)=k_0e^\delta,
  \qquad
  k_y(\delta)=k_0e^{-\delta}.
\end{equation}
Equation~\eqref{eq:supp-inplane-exchange} makes $H(\delta)$ an even function,
so
\begin{equation}
  \left.\frac{\dd H}{\dd\delta}\right|_{\delta=0}=0,
  \qquad
  H(\delta)-H(0)=\mathcal O(\delta^2).
  \label{eq:supp-isotropic-first-order-null}
\end{equation}
The local rank drop at $k_x=k_y$ follows from this symmetry, but it
does not imply that finite anisotropy has no effect.

An elliptical beam with a registered azimuth or noncollinear offset scans
can break this symmetry by supplying directional weighting. Independent
crystallographic information can instead assign the conductivity axes.
Numerically, a global check should also launch the optimizer from
both permutations, profile over an in-plane rotation angle, or compare the raw
forward traces directly.

\subsection{Two faces can be locally rank six but globally twofold}

Write the global tensor as
\begin{equation}
  \K=
  \begin{bmatrix}
    a&d&e\\ d&b&f\\ e&f&c
  \end{bmatrix}\succ0.
\end{equation}
Suppose measurements on the orthogonal $z$ and $x$ faces resolve the
cross-plane entries and the complete signed in-plane Schur tensors. Define the
inverse-tensor entries $r_{ij}=(\K^{-1})_{ij}$. The two inverse Schur
matrices supply $(r_{xx},r_{xy},r_{yy},r_{yz},r_{zz})$, while cross-plane
conductivity fixes $\Delta=\det(\K^{-1})$ through
$\Delta=(r_{yy}r_{zz}-r_{yz}^2)/a=(r_{xx}r_{yy}-r_{xy}^2)/c$.
These are six generically independent observable combinations, but the
missing $r_{xz}$ satisfies a quadratic with two roots. Thus their count
does not establish unique recovery of the six conductivity entries.

To construct the two tensors explicitly, define the
two measured off-diagonal Schur combinations
\begin{equation}
  D=d-\frac{ef}{c},
  \qquad
  R=f-\frac{de}{a},
  \qquad
  \lambda=1-\frac{e^2}{ac}>0.
\end{equation}
Here $D=S_{z,xy}$, $R=S_{x,yz}$, and
$S_{z,yy}=b-f^2/c$ are entries of the two measured face Schur tensors.
The two face data determine $e^2$, rather than the sign branch of $e$.  For
either $e=+|e|$ or $e=-|e|$, the remaining entries are reconstructed as
\begin{equation}
  d=\frac{D+(e/c)R}{\lambda},
  \qquad
  f=\frac{R+(e/a)D}{\lambda},
  \qquad
  b=S_{z,yy}+\frac{f^2}{c}.
  \label{eq:supp-two-face-branches}
\end{equation}
For invariant sets arising from a full anisotropic thermal-conductivity tensor,
both branches give physically admissible tensors that reproduce the same
two face-invariant sets. When $e\ne0$ they are
separated solutions and the local invariant
Jacobian can have rank six; when $e=0$ the branches merge and the local rank
drops to five.  This is a discrete two-face ambiguity, not the continuous
two-dimensional one-face family at fixed $(k_{zz},\Smat)$.

Three labeled orthogonal faces remove the branch.  Their cross-plane conductivity data give
$a$, $b$, and $c$; their Schur diagonals give the squared magnitudes of
$d$, $e$, and $f$; and their signed Schur off-diagonals select the mutually
consistent signs. Equivalently, if each coated measurement resolves its
complete signed Schur matrix, the inverse Schur matrices supply the
three in-plane blocks of $\K^{-1}$ and collectively contain all six
resistivity entries. This is a statement about resolved face invariants;
it does not assume that an arbitrary normalized signal determines them.

These examples delimit what SVD establishes.  Singular values describe local
parameter combinations in a declared scaling, not individual tensor entries,
and an absolute cutoff changes under nonsingular column scaling or altered row
weighting.  A full nonlinear identifiability claim additionally requires an
exact invariance argument or global diagnostics such as multistart,
profile-likelihood, and direct forward-response comparison.

\FloatBarrier

\section{Numerical implementation and coated-sample study design}
\label{sec:supp-numerical-methods}

\subsection{Independent evaluations of the coated surface response}
\label{sec:evaluators}

The numerical comparisons use an isotropic transducer, a scalar thermal
boundary resistance, and an anisotropic substrate. One evaluator solves
the complete anisotropic characteristic equation and calculates the
cross-plane heat flux directly; the other forms the substrate Schur
matrix. Both propagate the temperature--heat-flux state through the
same coating and interface. Their agreement tests the reduction without
assuming it in both evaluations. The finite substrate has an adiabatic
rear boundary.

The reference tensor is
\begin{equation}
  \K_0=\begin{bmatrix}
  10.4&2.0&-4.0\\2.0&7.0&4.0\\-4.0&4.0&8.6
  \end{bmatrix}\,\WmK,
  \label{eq:supp-counterexample-tensor}
\end{equation}
with $\Smat_0=\left[\begin{smallmatrix}8.5&3.9\\3.9&5.1
\end{smallmatrix}\right]\,\WmK$ and
$\K_{\mathrm{eq},0}=\diag(\Smat_0,8.6)$ in block notation.
Displayed entries are rounded; calculations use the exact Schur matrix.

The single-surface FDTR, spatial-scan, surface-map, and TDTR equivalence
comparisons use a \qty{100}{\nano\meter} isotropic transducer with
$k=\qty{100}{\watt\per\meter\per\kelvin}$ and
$C_v=\qty{2.4e6}{\joule\per\meter\cubed\per\kelvin}$, an interface
resistance $R=\qty{2e-8}{\meter\squared\kelvin\per\watt}$, and a
\qty{1.8}{\milli\meter} substrate with
$C_v=\qty{1.95e6}{\joule\per\meter\cubed\per\kelvin}$.
The effective Gaussian spatial radius is \qty{40}{\micro\meter}.
The line scan has 161 positions over $\pm\qty{160}{\micro\meter}$
at \qty{10}{\hertz}; the frequency sweep has 161 logarithmically
spaced frequencies from \qty{10}{\hertz} to \qty{1}{\mega\hertz}
at offset $(40,40)\,\unit{\micro\meter}$; and the
$61\times61$ surface map spans $\pm\qty{120}{\micro\meter}$.
These are equivalence controls with fixed stack properties. The
Au-coated reconstruction below uses its separately stated stack and
\qty{1}{\kilo\hertz}--\qty{10}{\mega\hertz} frequency range.

For responses $H$ and $H'$, we use the peak-scaled complex residual
$\varepsilon=\max_i|H_i-H_i'|/\max_i|H_i'|$ and phase discrepancy
$\varepsilon_\phi=\max_i|\phi_i-\phi_i'|$ in degrees. The coated
frequency sweep, line scan, and map each give
$\varepsilon<6\times10^{-16}$ and
$\varepsilon_\phi<7\times10^{-14}$ degrees. The phase-map residuals
are restricted to pixels exceeding \qty{1}{\percent} of the peak
amplitude. Supplementary Note~S3 describes the independent full-root
TDTR pulse-train calculation and its harmonic-truncation check.

\subsection{Coated component and conditional sensitivity checks}
\label{sec:jacobian-rank}
\label{sec:supp-validation-audit}

The componentwise calculation uses the same
\qty{100}{\nano\meter} coating and interface, with 81 positions and
81 frequencies over the ranges above. Its phase is the ordinary
unwrapped argument of the complex signal. All six physical tensor
entries are varied, one at a time, and the logarithmic phase derivatives
are assembled into spatial, frequency, and combined sensitivity matrices.
The optional joint calculation appends derivatives of the normalized
log amplitude $\ln|H_i| - \ln|H_0|$, where $H_0$ is the zero-offset
response at the same frequency. This amplitude normalization does not
change the definition of the phase channel.

All three phase matrices have rank four. For the combined scan, the
last two singular values, divided by the largest, are
$1.1\times10^{-12}$ and $1.0\times10^{-12}$; appending the amplitude
channel leaves them at the same numerical level. The individual
$k_{xz}$ and $k_{yz}$ sensitivities are nonzero, whereas the two
analytical directions that preserve $(k_{zz},\Smat)$ have residual
sensitivity below $5.2\times10^{-12}$. Thus the rank deficiency concerns
jointly varying tensor combinations, not vanishing individual columns.

The separate conditional comparison in Supplementary Note~S4 holds
$(k_{xx},k_{xy},k_{xz})$ fixed and fits only
$(k_{yy},k_{zz},k_{yz})$. Its spatial and frequency matrices each have
three columns and rank three; this is not a rank-three claim for the
six-unknown coated FDTR problem. The maximum Richardson changes for
its two nominal tensors are $2.3\times10^{-6}$ and $8.3\times10^{-5}$.

\subsection{Single-face ordinary-phase FDTR audit}
\label{sec:supp-single-face-fdtr-rank}

The $z$ face of the Au-coated reconstruction provides an independent
check using standard phase-versus-frequency data without zero-offset
subtraction. Three offset azimuths and 31 frequencies give 93 phase
values. Differentiation with respect to all six entries
$(k_{xx},k_{yy},k_{zz},k_{xy},k_{xz},k_{yz})$ gives a $93\times6$
matrix. With the physical column scales and noise weights specified
below, its four nonzero normalized singular values are
$(1,0.598,0.437,0.222)$; the remaining two are of order $10^{-16}$.
Its numerical rank is four using the relative threshold $10^{-8}$.
Refining the polar quadrature from $80\times64$ to $112\times80$
points leaves this result unchanged; two centered finite-difference
steps agree with the analytical Jacobian within $1.0\times10^{-8}$
in relative norm.

If $k_{xz}=k_{yz}=0$ is instead held fixed, the four remaining
parameters are $(k_{xx},k_{yy},k_{zz},k_{xy})$, and their
$93\times4$ matrix has normalized singular values
$(1,0.832,0.381,0.302)$ and rank four. For the unrestricted tensor,
the four resolved combinations are
$(k_{zz},S_{xx},S_{yy},S_{xy})$. These two parameter choices must be
distinguished when interpreting a reported four-component fit.

\subsection{Two-surface FDTR, SDTR, and TDTR equivalence}
\label{sec:supp-two-surface-comparison}

The comparison in \cref{sec:two-face-equivalence,fig:two-surface-equivalence}
uses the $z$ and $y$ faces, with right-handed local frames $(x,y,z)$ and
$(z,x,y)$. Both faces use the \qty{80}{\nano\meter} Au coating,
interface conductance, substrate, and \qty{40}{\micro\meter} pump and
probe radii specified in \cref{sec:supp-three-face-recovery}. The data
are noiseless, and phase is referenced to the modulated heating without
zero-offset subtraction.

Each FDTR sweep contains 161 logarithmically spaced frequencies from
\qty{1}{\kilo\hertz} to \qty{10}{\mega\hertz} at a
\qty{40}{\micro\meter} offset and local azimuth $45^\circ$.
The SDTR scan uses 161 signed offsets from $-4w$ to $4w$ along the
same azimuth at \qty{1}{\kilo\hertz}, with $w=\qty{40}{\micro\meter}$.
The centered TDTR calculation uses \qty{10}{\mega\hertz} modulation,
\qty{80}{\mega\hertz} repetition, \qty{2}{\pico\second} Gaussian
pump and probe full widths at half maximum, and 161 delays from
\qty{0.10}{\nano\second} to \qty{5.0}{\nano\second}.

Each tensor is evaluated independently using its full characteristic
roots and complete cross-plane heat flux. Original and equivalent
responses share the original-response peak normalization for each
method and face. Their maximum peak-scaled complex difference is below
$4\times10^{-16}$. Refining the FDTR/SDTR polar grid from
$80\times96$ to $112\times144$ nodes and the TDTR Gauss--Hermite
order from 24 to 32 changes the responses by less than
$5\times10^{-15}$ in the same metric. The TDTR reconstruction retains
harmonics $-16384\le\ell\le16384$; doubling the cutoff from 8192
changes the trace by less than $4\times10^{-8}$. This truncation check
is separate from the tensor-equivalence residual.

\subsection{Two-face recovery with a known zero coupling}
\label{sec:supp-two-face-known-zero}

The test in \cref{fig:two-face-known-zero-recovery} sets $k_{yz}=k_{zy}=0$
in $\K_0$ and fits $(k_{xx},k_{yy},k_{zz},k_{xy},k_{xz})$ on the $z,y$
faces. It uses the Au stack, 31 frequencies, and three offset azimuths
specified below for the three-face test. Stacking the directional scans
on the $z$ face followed by the $y$ face gives 186 ordinary phase values.
Independent uniform relative noise in $[-0.005,0.005]$ uses seed 20260912;
its standard deviations and physical-entry scales follow
\cref{eq:supp-three-face-noise,eq:supp-three-face-jacobian-svd} for the
five fitted entries.

The optimizer keeps the zero coupling exact by writing
$\K=\left[\begin{smallmatrix}a&d&e\\d&b&0\\e&0&c\end{smallmatrix}\right]$
with $a=t+d^2/b+e^2/c$, $t,b,c>0$, and signed $d,e$. This spans the
physically admissible tensors satisfying the known-zero constraint.
The analytic $186\times5$ physical-entry sensitivity matrix has normalized
singular values $(1,0.837,0.716,0.419,0.249)$ and rank five.
Its derivatives agree with a Richardson-extrapolated central difference
to $5.1\times10^{-10}$ relative error.

Six starts, comprising $10\mathbf I\,\WmK$ and five random admissible
tensors (seed 20260913), converge within $1.2\times10^{-8}$ relative
Frobenius distance. The fitted tensor error is \qty{1.8}{\percent};
the relative errors in the five entries are
$(1.69,0.26,0.39,5.96,2.88)\,\%$. Generation and fitting use
$160\times112$ and $80\times64$ polar grids, respectively. Refinement
to $224\times160$ changes phase by less than
$1.2\times10^{-12}$ degrees, and noiseless recovery gives a relative
tensor error below $1.5\times10^{-14}$. These checks test the declared
known-zero model; they do not infer the zero coupling from the data.

\subsection{Three-face noisy phase reconstruction}
\label{sec:supp-three-face-recovery}

The example in \cref{sec:three-face-noisy-recovery} uses the exact
$\K_0$ of \cref{eq:counterexample-tensor}. Each face is evaluated as a planar
stack away from edges, with cyclic right-handed frames $(x,y,z)$,
$(y,z,x)$, and $(z,x,y)$. The substrate is \qty{1.8}{\milli\meter} thick
with an adiabatic rear boundary and
$C_v=\qty{1.95e6}{\joule\per\meter\cubed\per\kelvin}$.
The isotropic Au transducer has thickness \qty{80}{\nano\meter},
conductivity \qty{180}{\watt\per\meter\per\kelvin}, and room-temperature
volumetric heat capacity
\qty{2.49e6}{\joule\per\meter\cubed\per\kelvin}. The heat-capacity value
follows Table~1 of Tang and Dames~\cite{Tang2021}. The thermal boundary conductance
is $G=\qty{1e8}{\watt\per\meter\squared\per\kelvin}$, equivalent to
$R=G^{-1}=\qty{1e-8}{\meter\squared\kelvin\per\watt}$. Both Gaussian $1/e^2$
beam radii and the offset are \qty{40}{\micro\meter}. Each face uses
31 logarithmically spaced frequencies from \qty{1}{\kilo\hertz} to
\qty{10}{\mega\hertz} at offset azimuths $0^\circ$, $45^\circ$, and
$90^\circ$, giving 279 phase observations. For each face
$n\in\{z,x,y\}$, let $\boldsymbol\phi_{n,\alpha}\in\mathbb R^{31}$
contain the phase values at azimuth $\alpha$ in increasing frequency order.
The combined data vector is
\begin{equation}
  \begin{aligned}
    \boldsymbol\phi_n
      &=(\boldsymbol\phi_{n,0^\circ}^{\mathsf T},
         \boldsymbol\phi_{n,45^\circ}^{\mathsf T},
         \boldsymbol\phi_{n,90^\circ}^{\mathsf T})^{\mathsf T}
         \in\mathbb R^{93},\\
    \boldsymbol\phi
      &=(\boldsymbol\phi_z^{\mathsf T},
         \boldsymbol\phi_x^{\mathsf T},
         \boldsymbol\phi_y^{\mathsf T})^{\mathsf T}
         \in\mathbb R^{279}.
  \end{aligned}
  \label{eq:supp-three-face-data-vector}
\end{equation}

The signal is the unwrapped argument of the complex lock-in response,
relative to the modulated heating, without zero-offset subtraction.
For each observation $m$, the synthetic data
and fixed noise standard deviation are
\begin{equation}
  \phi_m^{\mathrm{obs}}=\phi_m(\K_0)(1+\eta_m),\qquad
  \eta_m\sim\mathcal U(-0.005,0.005),\qquad
  s_m=\frac{0.005|\phi_m(\K_0)|}{\sqrt{3}}.
  \label{eq:supp-three-face-noise}
\end{equation}
The independent noise realization uses seed 20260910. The largest sampled
relative perturbation is below \qty{0.5}{\percent}. The weights use the
known synthetic noise law, not an estimated experimental variance.
We minimize
\begin{equation}
  \sum_m\left[\frac{\phi_m(\K)-\phi_m^{\mathrm{obs}}}{s_m}\right]^2
  \label{eq:supp-three-face-nlls}
\end{equation}
over six free coordinates of
$\K=(10\,\WmK)\mathbf L\mathbf L^{\mathsf T}$, where $\mathbf L$ is lower
triangular with exponentiated diagonal entries. All six tensor entries are
therefore fitted without sign constraints on the couplings. SciPy
\texttt{least\_squares} uses an analytic phase Jacobian, function and step
tolerances of $10^{-11}$, and gradient tolerance $10^{-9}$. Eight fixed starts comprise $10\mathbf I\,\WmK$ and seven
random full anisotropic thermal-conductivity tensors with random eigenvectors and eigenvalues drawn
log-uniformly from 2 to $30\,\WmK$ (seed 20260911). All converge to the
same fit within $9.2\times10^{-9}$ relative Frobenius distance.

The six sensitivity parameters are
$\mathbf k=(k_{xx},k_{yy},k_{zz},k_{xy},k_{xz},k_{yz})^{\mathsf T}$.
The Cholesky coordinates are used only by the optimizer. Each coupling
derivative varies both symmetric matrix positions together. We form
\begin{equation}
  J_{mj}=\left.\frac{\partial\phi_m}{\partial k_j}\right|_{\K_0},
  \qquad
  \widetilde{\mathbf J}
    =\diag(s_m^{-1})\mathbf J\diag(a_j)
    =\mathbf U\boldsymbol\Sigma\mathbf V^{\mathsf T},
  \label{eq:supp-three-face-jacobian-svd}
\end{equation}
where $\mathbf J,\widetilde{\mathbf J}\in\mathbb R^{279\times6}$,
$a_{ii}=k_{ii,0}$, and $a_{ij}=\sqrt{k_{ii,0}k_{jj,0}}$.
These fixed column scales and row weights make the reported singular values
dimensionless. The six singular values of the three-face matrix are
$(549.3,313.3,284.6,233.4,169.9,101.9)$.
The two-face subset below uses the $z$ and $x$ faces, distinct from
the $z$--$y$ signal comparison in \cref{fig:two-surface-equivalence}.
At $\K_0$, the singular values normalized by the largest are
\begin{equation}
\begin{aligned}
 z &: (1,\;0.5981,\;0.4368,\;0.2222,\;<10^{-15},\;<10^{-15}),\\
 z,x &: (1,\;0.5664,\;0.4927,\;0.2900,\;0.1727,\;0.1590),\\
 z,x,y &: (1,\;0.5704,\;0.5181,\;0.4249,\;0.3093,\;0.1856).
\end{aligned}
\label{eq:supp-three-face-spectra}
\end{equation}
\begin{figure}[t]
  \centering
  \includegraphics[width=0.68\textwidth]{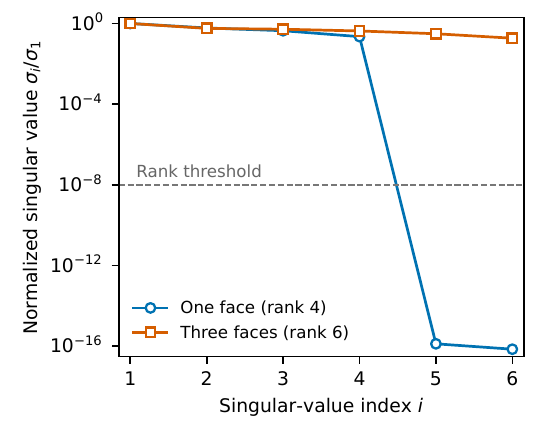}
  \caption{Normalized singular spectra for the one-face and three-face
  phase protocols at $\K_0$, with physical-entry scaling and noise weights
  defined in the text. The one-face matrix has two structural null
  directions; all six directions are resolved by the three-face matrix.
  The dashed line is the relative numerical rank threshold.}
  \label{fig:supp-three-face-sensitivity}
\end{figure}

\Cref{fig:supp-three-face-sensitivity} contrasts the one-face and three-face
spectra. The coated phase ranks for one, two, and three faces are four, six,
and six, respectively. These are six-parameter sensitivity matrices;
rank six at two faces is a local statement and does not remove the
global branch discussed below.
Fresh physical-entry derivatives reproduce the saved Jacobian; independent
central differences agree to $3.6\times10^{-9}$ relative error. The SVD
reconstructs the combined matrix to $2.5\times10^{-15}$ relative error.
The absolute relative entry errors, in the same order, are
$(0.320,0.092,0.049,6.625,1.583,0.232)\,\%$.
The phase residual RMS is \qty{0.128}{\degree}, and its noise-weighted
RMS is 0.994. Local linearized standard deviations from the inverse
information matrix are
$(0.033,0.026,0.031,0.073,0.048,0.044)\,\WmK$.
They include only the declared measurement noise; all auxiliary parameters
are exact.

The forward solver uses the full-flux planar-stack response integrated
over radial and angular Fourier coordinates. The generation grid has
$160\times112$ nodes and the fitting grid $80\times64$.
Their maximum phase difference is below
$1.1\times10^{-12}$ degrees at both truth and fit, far below the noise.
An additional $320\times224$ refinement changes the generated phase by
less than $1.8\times10^{-12}$ degrees.
A noiseless recovery using the separate generation grid has relative
tensor error $1.3\times10^{-14}$.
Finite-difference checks verify the analytic Jacobian to relative error
$1.3\times10^{-9}$, and the existing full-root implementation reproduces
the new complex response to within $4.7\times10^{-15}$ relative error.

For the same $z$--$x$ subset used in the sensitivity analysis, the
exact two-face branch in Supplementary Note~S5 supplies the control
\begin{equation}
  \K_{\mathrm{twin}}\simeq
  \begin{bmatrix}
    10.4&7.4&4.0\\
    7.4&11.9&7.6\\
    4.0&7.6&8.6
  \end{bmatrix}\,\WmK.
\end{equation}
Using unrounded entries, this full anisotropic thermal-conductivity tensor reproduces the $z$- and $x$-face
phase curves to numerical precision, whereas its $y$-face curve differs
by up to \qty{4.68}{\degree}. Thus even the locally full-rank two-face
protocol retains a global ambiguity.
The three-face fit demonstrates recovery for one tensor and one noise
realization under the same physical model used for generation. It does
not test correlated measurement error, model mismatch, uncertain inputs,
or imperfect face registration.

\bibliography{bib/references}

\end{document}